\documentclass{article}
\usepackage{amssymb}
\usepackage{graphicx}
\usepackage{amsmath}

\input{tcilatex}

\begin{document}

\begin{quotation}
{\Huge Towards physically motivated }

{\Huge proofs \ of \ the \ Poincar\'{e} \ and }

{\Huge geometrization conjectures \ \ \ \ \ \ \ \ \ \ \ \ \ \ \ \ \ \ \ \ \
\ \ \ \ \ \ \ \ \ \ \ \ \ \ \ \ \ \ \ \ \ \ \ \ \ \ \ \ \ \ \ \ \ \ \ \ \ \
\ \ \ \ \ \ \ \ \ \ \ \ \ \ \ \ \ \ \ \ \ \ }
\end{quotation}

\ \ \ \ \ \ \ \ \ \ \ \ \ \ \ \ \ \ \ \ \ \ \ \textbf{Arkady L.Kholodenko}

\ \ \ \ \ \ 

$\ \ \ \ \ $375 H.L.Hunter Laboratories, Clemson University, Clemson,

\ \ \ \ \ SC 29634-0973, USA. e-mail: string@clemson.edu

$\ \ \ \ \ \ \ $

\bigskip

Although the Poincar$e^{\prime }$ and the geometrization conjectures were
recently proved by Perelman, the proof relies heavily on properties of the
Ricci flow previously investigated in great detail by Hamilton. Physical
realization of such a flow can be found, \ for instance, in \ the work by
Friedan (Ann.Physics 163 (1985), 318-419). In his work the renormalization
group flow for a nonlinear sigma model in $2+\varepsilon $ dimensions was
obtained and studied. For $\varepsilon =0,$ by \ approximating the $\beta -$
function for such a flow by the\ lowest order terms in the sigma model
coupling constant, the equations for Ricci flow are obtained. In view of
such an approximation, the existence of this type of flow in nature is
questionable. In this work, we find totally independent justification for
the existence of \ Ricci flows in nature. This is achieved by developing a
new formalism extending the results of two dimensional \ conformal field
theories (CFT's) to three and higher dimensions. Equations describing
critical dynamics of these CFT's are examples of the Yamabe and Ricci flows
realizable in nature. Although in the original works by Perelman some
physically motivated arguments can be found, their role in his proof remain
rather obscure. In this paper, steps are made toward making these arguments
more explicit thus creating an opportunity for developing alternative, more
physically motivated, proofs of the Poincar$e^{\prime }$ and \
geometrization conjectures.

\ 



\pagebreak

\ \ \ \ \ \ 

\section{\protect\bigskip Introduction}

\ \ \ \ \ The history of physics is full of examples of situations when
experimental observations lead to very deep mathematical results. Beginning
with Newton's dynamics equations, which gave birth to analysis, through the
heat and wave equations, which gave birth to mathematical physics, through
Maxwell's equations (experimentally discovered by Faraday), which ultimately
had led to special relativity on one hand and the theory of homology and
cohomology on another, etc. thus stimulating \ the development of topology.
Methods of celestial mechanics, especially those developed by Poincar$%
e^{\prime }$, had led Bohr and his collaborators to quantum mechanics, on
one hand, and to KAM and chaos theory, on another. \ A simple observation by
Einstein that gravity can be locally eliminated in the freely falling
elevator cabin had led to general relativity, on one hand, and to great
advancements in differential geometry, on another. \ Based on these
examples, it is only natural to expect that the recent complete proof of the
Poincar$e^{\prime }$ and geometrization conjectures by Perelman [1-3] should
be connected with some processes taking place in the real world. The purpose
of this work is to demonstrate that this is indeed the case. Since this
paper is not a review, we made no efforts to list all the latest
physics-related results inspired by Perelman's works. Instead, we shall only
quote those works which are immediately relevant to the content of our paper.

We begin with the observation that in the work of Friedan [4] on the
nonlinear sigma model in $2+\varepsilon $ dimensions, published in 1985%
\footnote{%
It is based on his PhD thesis completed in August of 1980.}, the reference
was made to a talk by Bourguignon (delivered in 1979) "On Ricci curvature
and Einsten metric" in which for the first time the idea of usefulness of
Ricci flows to differential geometry and related disciplines was introduced,
e.g. see Ref.10 in Friedan's paper. This idea was picked up by Hamilton in
1982 [5] who began from that time on a systematic study of Ricci flows.
Although written with exceptional clarity, his results summarized in Ref.[6]
apparently escaped physicists attention for many years. This is caused by
several factors, which are worth mentioning. The Ricci-type flow equations
appeared in the physics literature in the 80's in connection with the
renormalization group analysis of nonlinear sigma models and also of strings
propagating on curved backgrounds [4,7] without any reference to the
parallel developments in mathematics. Because of this, the point of view was
developed in the physics literature, especially that related to string
theory [8], that all major results of general relativity along with quantum
corrections can be deduced from the action functional \ whose minimization
produces the Euler-Lagrange equations for the renormalization group\ (RG)
flow for strings in curved backgrounds, e.g. see [8], Ch-r 3, page 180. In
the simplest case such a functional ( e.g. see Eq.(3.4.58) on page 180) is a
modification of the Hilbert-Einsten functional describing gravity, leading
to the action functional for the dilaton gravity. The only problem with such
a functional lies in the fact that it is "living" in 26 dimensional
space-time for a bosonic string and in 10 dimensional space-time for a
fermionic (heterotic) string [9]. These facts, apparently were known to
Grisha Perelman who used them in his calculations, many of which (but not
those involving such functionals) are specific only for 3 dimensional \
Riemannian manifolds and, hence, are not readily extendable to higher
dimensions and a pseudo-Riemannian-type of metric.

At this point it is essential to notice that both in the work by Friedan [4]
and in the string-theoretic literature just cited, the dilaton gravity
functional emerges only as a result of the lowest order approximation in
RG-type calculations. This fact can be seen very clearly on page 322 of
Friedan's paper, where in his Eq.s (1.5) and (1.6) one has to put $%
\varepsilon =0$ and retain only terms of order $O(T^{0})$ in the sigma model
coupling constant $T^{-1}.$ Such a retention then straightforwardly leads to
the equation for the Ricci flow.\footnote{%
At the same time, to obtain the generating functional whose minimization
produces the Euler-Lagrange equaton for such a flow is nontrivial. In
mathematics literature the credit for finding such a functional goes to
Perelman. In physics literature this functional, Eq.(3.4.58) on page 180 of
Ref.[8], was known much earlier. It describes the dilaton gravity (albeit in
the nonphysical space-time dimensions).} When terms of order $T$ and higher
are included into these RG flow equations, such an RG flow is no longer of
Ricci type. The procedure of an ad hoc truncation of an infinite asymptotic
RG series expansion for the RG $\beta -$ function (e.g. see Eq.(1.5) in
Friedan's paper) \ in order to get the needed equation for the Ricci flow
makes the physical existence of such a flow doubtful. Since both the Poincar$%
e^{\prime }$ and the geometrization conjectures were proved with help of the 
\textit{unperturbed }(\textit{nonmodified by terms other than those induced
by diffeomorphisms}) Ricci flows [1-3], it is essential to find physical
processes that employ such unperturbed Ricci-type flows. The present work
provides examples of such physical processes. It enables us to reconnect
recent significant mathematical advances with physical processes realizable
in nature in accord with the logic of development of mathematical physics
described at the beginning of this section.

In view of space limitation, before discussing the content of this paper in
some detail, we would like to mention the following. The existing proofs of
the Poincar$e^{\prime }$ and the geometrization conjectures are either too
compressed [1] or too bulky [2,3,10]. In both cases, it may take a
considerable amount of time for the nonexpert reader to learn and to
understand these papers. Fortunately, there are succinct publications the
present the key ideas of the proof in a language that is considerably more
familiar to a much wider group of people with mathematical background. \
Among these reviews, we would especially recommend those written by
Bessieres [11], on the proof of the Poincar$e^{\prime }$ conjecture, and by
Kapovich [12] on the proof of the geometrization conjecture. \ This paper is
written under the assumption that our readers have previously read these
reviews.

The organization of the rest of this paper is the following. In Section 2
the scaling analysis of the Ginzburg-Landau (G-L) functional, widely used in
the theory of critical phenomena, is discussed using both arguments known in
physics literature and those known in mathematics. In particular, we argue
that in the physics literature there is a marked difference in treatments of
two and the higher dimensional models exhibiting critical behavior. While in
two dimensions the full conformal invariance is used (at least at
criticality), in dimensions higher than two only scale invariance is taken
into account at criticality. Using scaling arguments from both physics and
mathematics, we demonstrate how the existing scaling treatments in higher
dimensions can be improved in order to take into account the effects of full
conformal invariance. This improvement is brought to completion in Section 3
where we connect the G-L functional used in the physics literature with the
Yamabe functional known in mathematics. We demonstrate that, if properly
interpreted, both functionals upon minimization produce the same G-L-type
equations used in literature on critical phenomena. The advantage of working
with Yamabe functionals lies in our observations that a) such a functional
can be rewritten in terms of the Hilbert-Einstein action functional for
Euclidean gravity and, b) that such a functional is manifestly conformally
invariant. Such a connection between the H-E action and its
pseudo-Riemannian extension is needed in applications in relativity \ which
are studied in Section 4. Connections between the G-L and Yamabe functionals
exist only for dimensionality higher than two. In two dimensions one has to
find an analog of the G-L-Yamabe functional. This task is accomplished in
Section 5 in which not only such an analog is found but, also, its
connection with known results in string and CFT. Specifically, we were able
to connect such an analog with the much earlier results obtained by Distler,
Kawai and David for 2d quantum gravity in noncritical dimensions
pedagogically summarized in the book by Hatfield [13]. Applications of their
results to CFT are summarized in the lecture notes by Abdalla et al [14]
and, more recently, in the review by Nakayama [15]. Because of such
connections, the extension of these 2 dimensional \ CFT results to higher
dimensions can be done quite systematically. This is discussed in Section 6.
In Section 7 we develop an alternative method allowing us to reobtain
results known already for 2 and higher dimensions. This formalism is based
on the observation that results obtained in previous sections are just
static solutions of more general equations describing critical dynamics for
systems of the G-L-type. By noticing that mathematically such dynamics
coincides with that known for the Yamabe-type flows discovered by Hamilton
[16], we follow the logic of his works in order to relate the Yamabe flow to
the more general Ricci flow of which the Yamabe flow is a special case. To
do so we use some results by Perelman [1], which he was using in his proofs
of the Poincar$e^{\prime }$ and the geometrization conjectures. Our use of
Perelman's results is not mechanical however. By exposing physical arguments
hidden in his work, we were able to reobtain many of his results \ much
simpler \ as compared \ to treatments which can be found in current works by
mathematicians [2,3,10] aimed at explaining and elaborating on many hidden
details of Perelman's original proof. Finally, in Section 8 we discuss other
possible physical systems whose dynamics is expected to be described in
terms of Ricci flows. We also discuss the conditions under which experiments
on critical dynamics should be conducted in order to detect the effects of
Ricci and Yamabe flows.

\section{Scaling analysis of the G-L functional}

Conventional scaling analysis of the G-L functional used routinely in
physics literature can be found, for example, in Ref.[17]. This analysis
differs somewhat from that for the $\phi ^{4}$ model as described in the
monograph by Itzykson and Zuber [18]. For the sake of the discussion that
will follow, we would like to provide a sketch of the arguments for both
cases now.

We begin with the $\phi ^{4}$ model following Ref.[18]. Let $\mathcal{L}(x)$
be the Lagrangian of this scalar field model whose action functional in $d$
dimensions $S[\phi ]$ is given by $S[\phi ]=\int d^{d}x\mathcal{L}(x)$. Let
furthermore $\lambda $ be some nonnegative parameter. Then the requirement
that $S[\phi ]$ be independent of $\lambda $, i.e. $\int d^{d}x\mathcal{L}%
(x)=\int d^{d}x\lambda ^{d}\mathcal{L}(\lambda x),$ leads to the constraint 
\begin{equation}
\int d^{d}x(x\cdot \frac{\partial }{\partial x}+d)\mathcal{L}(x)=0  \tag{2.1}
\end{equation}%
obtained by differentiation \ of \ $S[\phi ]$ with respect to $\lambda $
with $\lambda $ being equalled to one at the end of calculation. For $%
\mathcal{L}(x)$ given by 
\begin{equation}
\mathcal{L}(x)=\frac{1}{2}\left( \bigtriangledown \phi \right) ^{2}+\frac{%
m^{2}}{2}\phi ^{2}+\frac{\hat{G}}{4!}\phi ^{4}  \tag{2.2}
\end{equation}%
the change of $\mathcal{L}(x)$ under the infinitesimal scale transformation
is given by\footnote{%
In arriving at this result we took into account Eq.(13-40) of Ref.[18] along
with condition $D=\frac{d}{2}-1$, with $D$ defined in Eq.(13-38) of the same
reference. Also, we changed signs (as compared to the original source) in $%
\mathcal{L}(x)$ to be in accord with the accepted conventions for the G-L
functional.} 
\begin{equation}
\frac{\delta \mathcal{L}}{\delta \varepsilon }=(x\cdot \frac{\partial }{%
\partial x}+d)\mathcal{L}(x)\text{ }\mathcal{+}(d-4)\frac{\hat{G}}{4!}\phi
^{4}-m^{2}\phi ^{2}.  \tag{2.3}
\end{equation}%
Comparison between Eq.s(2.1) and (2.3) implies that the action $S[\phi ]$ is
scale invariant if $d=4$ and $m^{2}=0.$ The result just obtained raises an
immediate question: given that the massless G-L action is scale invariant
for $d=4$, will it also be conformally invariant under the same conditions?
We provide the answer to this question in several steps.

First, let $M$ be some Riemannian manifold whose metric is $g$. Then, any
metric $\tilde{g}$ conformal to $g$ can be written as $\tilde{g}=e^{f}g$
with $f$ being a smooth real valued function on $M$ [19]. Let $\Delta _{g}$
be the Laplacian associated with metric $g\footnote{%
That is $\Delta _{g}\Psi $ $=-(\det g)^{-\frac{1}{2}}\partial
_{i}(g^{ij}(\det g)^{\frac{1}{2}}\partial _{j}\Psi )$ for some scalar
function $\Psi (x).$}$ and, accordingly, let $\Delta _{\tilde{g}}$ be the
Laplacian associated with metric $\tilde{g}.$ Richardson [20] demonstrated
that 
\begin{equation}
\Delta _{\tilde{g}}=e^{-f}\Delta _{g}-\frac{1}{2}(\frac{d}{2}%
-1)fe^{-f}\Delta _{g}-\frac{1}{2}(\frac{d}{2}-1)e^{-f}(\Delta _{g}f)+\frac{1%
}{2}(\frac{d}{2}-1)e^{-f}\Delta _{g}\circ f  \tag{2.4}
\end{equation}%
(here $\left( \Delta _{g}\circ f\right) \Psi $ should be understood as $%
\Delta _{g}(f(x)\Psi (x)).$ \ Eq.s(2.3) and (2.4) imply that the conformally
invariant (string-type) functional $S[\mathbf{X}]=\int_{M}d^{2}x\sqrt{g}%
(\bigtriangledown _{g}\mathbf{X})\cdot (\bigtriangledown _{g}\mathbf{X})$
exists only for $d=2$. For $d>2$,\ in view of Eq.(2.4), the conformal
invariance of the $\phi ^{4}$ model is destroyed. Thus, using conventional
field-theoretic perturbational methods one encounters a problem with
conformal invariance at the zeroth order level (in the coupling constant $%
\hat{G}$), unless $d=2$. This problem formally does not occur if \ one
requires our physical model to be only scale invariant. This is undesirable
however in view of the fact that in 2 dimensions an arbitrary conformal
transformation of the metric tensor $g$ of the underlying two dimensional
manifold $M$ is permissible at criticality [21]. Abandoning the requirement
of general conformal invariance in 2 dimensions in favor of scale invariance
in higher dimensions would destroy all known string-theoretic methods of
obtaining exact results in two dimensions. Our general understanding of
critical phenomena depends crucially on our ability to solve two dimensional
models exactly. All critical properties for the same type of models in
higher dimensions are expected to hold even in the absence of exact
solvability. Fortunately, the situation can be considerably improved by
reanalyzing and properly reinterpreting the scaling results for dimensions
higher than 2.

This \ observation leads to the next step in our arguments. Using the book
by Amit [17], we consider first the scaling of the non interacting ( free )
G-L theory whose action functional is given by 
\begin{equation}
S[\phi ]=\int d^{d}x\{(\bigtriangledown \phi )^{2}+m^{2}\phi ^{2}\}. 
\tag{2.5}
\end{equation}%
Suppose now that, upon rescaling, the field $\phi $ transforms according to
the rule\footnote{%
Notice that this is already a special kind of conformal transformation}: 
\begin{equation}
\tilde{\phi}(Lx)=L^{\omega }\phi (x).  \tag{2.6}
\end{equation}%
If we require 
\begin{equation}
\int d^{d}x\{(\bigtriangledown \phi )^{2}+m^{2}\phi ^{2}\}=\int
d^{d}xL^{d}\{(\tilde{\bigtriangledown}\tilde{\phi})^{2}+\tilde{m}^{2}\tilde{%
\phi}^{2}\}  \tag{2.7}
\end{equation}%
and use Eq.(2.6) we obtain, 
\begin{equation}
S[\phi ]=\int d^{d}xL^{d}\{(\bigtriangledown \phi )^{2}L^{2\omega -2}+\tilde{%
m}^{2}\phi ^{2}L^{2\omega }\}.  \tag{2.8}
\end{equation}%
In order for the functional $S[\phi ]$ to be scale invariant the mass $m^{2}$
should scale as follows: $\tilde{m}^{2}=m^{2}L^{-2}$. With this requirement
the exponent $\omega $ is found to be: $\omega =1-\frac{d}{2}.$

\ The above scaling can be done a bit differently following the same
Ref.[17]. For this purpose we notice that although the action $S[\phi ]$ is
scale invariant, there is some freedom of choice for the dimensionality of
the field $\phi $. For instance, instead of $S[\phi ]$ we can consider 
\begin{equation}
S[\phi ]=\frac{1}{a^{d}}\int d^{d}x\{(\bigtriangledown \phi )^{2}+m^{2}\phi
^{2}\}  \tag{2.9}
\end{equation}%
with $a^{d}$ being some volume, say, $a^{d}=\int d^{d}x$ . Then, by
repeating arguments related to Eq.(2.8), we obtain: $\omega =1$ (instead of $%
\omega =1-\frac{d}{2}$ previously obtained $)$ in accord with Eq.(2-66) of \
the book \ by Amit [17]$.$ Although, from the point of view of scaling
analysis both results are actually equivalent, they become quite different
if we want to extend such scaling analysis by considering general conformal
transformations. Even though, in view of Eq.(2.4), such a task seems
impossible to accomplish, fortunately, this is not true as we would like to
demonstrate. This leads us to the next step.

Noticing that the mass term scales as the scalar curvature $R$ \ for some
Riemannian manifold $M$, i.e. the scaling $\tilde{m}^{2}=m^{2}L^{-2}$ is 
\textit{exactly the same} as the scaling of $R$ given by 
\begin{equation}
\tilde{R}=L^{-2}R.  \tag{2.10}
\end{equation}%
This result can be found, for example, in the book by Wald, Ref. [22],
Eq.(D.9), page 446. In general, the scalar curvature $R(g)$ changes under
the conformal transformation $\hat{g}=e^{2f}g$ according to the rule [19] 
\begin{equation}
\hat{R}(\hat{g})=e^{-2f}\{R(g)-2(d-1)\Delta _{g}f-(d-1)(d-2)\left\vert
\bigtriangledown _{g}f\right\vert ^{2}\}  \tag{2.11}
\end{equation}%
where $\Delta _{g}f$ is the Laplacian of $f$ and $\bigtriangledown _{g}f$ is
the covariant derivative defined with respect to the metric $g$. \ From here
we see that, indeed, for constant $f$ 's the scaling takes place in accord
with Eq.(2.10). Now, however, we can do more. Following Lee and Parker [19],
we simplify the above expression for $R$. For this purpose we introduce a
substitution: $e^{2f}=\varphi ^{p-2}$, where $p=\frac{2d}{d-2},$ so that $%
\hat{g}=\varphi ^{p-2}g.$ With such a substitution, Eq.(2.11) acquires the
following form: 
\begin{equation}
\hat{R}(\hat{g})=\varphi ^{1-p}(\alpha \Delta _{g}\varphi +R(g)\varphi ), 
\tag{2.12}
\end{equation}%
with $\alpha =4\frac{d-1}{d-2}.$ Clearly, such an expression makes sense
only for $d\geq 3$ and breaks down for $d=2$. But we know already the action 
$S[\mathbf{X}]$ which is both scale and conformally invariant in $d=2$. It
is given after Eq.(2.4). Fortunately, the results obtained in this section \
can be systematically extended in order to obtain actions, which are both
scale and conformally invariant in 3 dimensions and higher. This is
described in the next section.

\bigskip

\section{G-L functional and the Yamabe problem}

\bigskip

We begin with the following observation. Let $\tilde{R}(\tilde{g})$ in
Eq.(2.11) be some constant (that this is indeed the case, we shall
demonstrate shortly). Then Eq.(2.12) acquires the following form 
\begin{equation}
\alpha \Delta _{g}\varphi +R(g)\varphi =\hat{R}(\hat{g})\varphi ^{p-1}. 
\tag{3.1}
\end{equation}%
By noticing that $p-1=\frac{d+2}{d-2}$ we obtain at once : $p-1=3$ (for $d=4$%
) and $p-1=5$ ( for $d=3$). These are familiar Ginzburg-Landau values \ for
the exponents of \ interaction terms for critical and tricritical G-L
theories [23]. Once we recognize these facts, the action functional
producing the G-L-type Eq.(3.1) can be readily constructed. For this purpose
it is sufficient to rewrite Eq.(2.9) in a manifestly covariant form. We
obtain, 
\begin{equation}
S[\varphi ]=\frac{1}{\left( \int_{M}d^{d}x\sqrt{g}\varphi ^{p}\right) ^{%
\frac{2}{p}}}\int_{M}d^{d}x\sqrt{g}\{\alpha (\bigtriangledown _{g}\varphi
)^{2}+R(g)\varphi ^{2}\}\equiv \frac{E[\varphi ]}{\left\Vert \varphi
\right\Vert _{p}^{2}}.  \tag{3.2}
\end{equation}%
Minimization of this functional produces the following Euler-Lagrange
equation 
\begin{equation}
\alpha \Delta _{g}\bar{\varphi}+R(g)\bar{\varphi}-\lambda \bar{\varphi}%
^{p-1}=0  \tag{3.3}
\end{equation}%
with constant $\lambda $ denoting the extremum value for the ratio: 
\begin{equation}
\lambda =\frac{E[\bar{\varphi}]}{\left\Vert \bar{\varphi}\right\Vert _{p}^{p}%
}=\inf \{S[\varphi ]:\hat{g}\text{ conformal to }g\text{\}.}  \tag{3.4}
\end{equation}%
In accord with the Landau theory of phase transitions [24] it is expected
that the conformal factor $\varphi $ is a smooth nonnegative function on $M$
achieving its extremum value $\bar{\varphi}$ . Comparison between Eq.s(3.1)
and (3.3) implies that $\lambda =\tilde{R}(\tilde{g})$ as required. These
results belong to Yamabe, who obtained the\ Euler-Lagrange G-L-type Eq.(3.3)
upon minimization of the functional $S[\varphi ]$ without any knowledge of
the Landau theory$.$ The constant $\lambda $ is known in literature as the 
\textit{Yamabe invariant }[19,25]. Its value is an invariant of the
conformal class $(M,g)$. Given these facts, we obtain the following:

\ 

\textbf{Definition 3.1}. The \textit{Yamabe problem} lies in finding a
compact Riemannian manifold $(\mathit{M,g})$ of dimension $n\geq 3$\ whose
metric \ is conformal to the metric $\hat{g}$\ of a constant scalar
curvature.

\ 

Subsequent developments, e.g. that given in Ref.[26,27], extended this
problem to manifolds with boundaries and to non- compact manifolds. It is
not too difficult to prove that the (Yamabe-Ginzburg-Landau-like) functional
is manifestly conformally invariant. For this purpose, we need to rewrite
Eq.(2.12) in the following equivalent form 
\begin{equation}
\varphi ^{p}\hat{R}(\hat{g})=(\alpha \varphi \Delta _{g}\varphi +R(g)\varphi
^{2}).  \tag{3.5}
\end{equation}%
This can be used in order to rewrite $E[\varphi ]$ as follows : $E[\varphi
]=\int d^{d}x\sqrt{g}\varphi ^{p}\hat{R}(\hat{g})$. Next, by noting that $%
\int d^{d}x\sqrt{\hat{g}}=\int d^{d}x\sqrt{g}\varphi ^{p},$ we can rewrite
the Yamabe functional in the Hilbert-Einstein form 
\begin{equation}
S[\varphi ]=\frac{\int d^{d}x\sqrt{\hat{g}}\hat{R}(\hat{g})}{\left( \int
d^{d}x\sqrt{\hat{g}}\right) ^{\frac{2}{p}}}  \tag{3.6}
\end{equation}%
where both the numerator and the denominator are invariant with respect to
conformal changes in the metric, e.g. scale change.

\ 

\textbf{Remark 3.2}. In order to use these results in statistical mechanics,
we have to demonstrate that the extremum of the Yamabe functional $S[\varphi
]$ is realized for manifolds $M$ whose scalar curvature $R(g)$ in Eq.(3.3)
is also constant. This is the essence of the Yamabe problem (as defined
above) for physical applications. Fortunately, the Yamabe problem has been
solved positively by several authors starting with Yamabe himself (whose
proof contained some mistakes however) and culminated in the work by Schoen
[28]. Details can be found in the review paper by Lee and Parker [19] and in
the monograph by Aubin [29].

\ 

\textbf{Remark 3}.\textbf{3}. In view of the relation $\int d^{d}x\sqrt{\hat{%
g}}=\int d^{d}x\sqrt{g}\varphi ^{p},$ it is clear that for the fixed
background metric $g$ Eq.(3.3) can be obtained alternatively using the
following variational functional 
\begin{equation}
\tilde{S}[\varphi ]=\int d^{d}x\sqrt{g}\{\alpha (\bigtriangledown
_{g}\varphi )^{2}+R(g)\varphi ^{2}\}-\tilde{\lambda}\int d^{d}x\sqrt{g}%
\varphi ^{p}  \tag{3.7}
\end{equation}%
where the Lagrange multiplier $\tilde{\lambda}$ \footnote{$\tilde{\lambda}%
=\lambda \frac{2}{p}\equiv \lambda \frac{d-2}{d}$}is responsible for the
volume constraint. Such a form of the functional $\tilde{S}[\varphi ]$
brings this higher dimensional result in accord with that to be developed
below for two dimensions (e.g. see Section 5, Eq.(5.24), and Section 7%
\textbf{\ }).

\ 

Apart from the normalizing denominator, Eq.(3.6) represents the
Hilbert-Einstein action for pure gravity defined for $\ $Riemannian $d$
dimensional space. The denominator, the volume $V$ taken to power $\frac{2}{p%
},$ serves the purpose of making $S[\varphi ]$ manifestly conformally
invariant, Ref.[28], page 150.

\section{\protect\bigskip From G-L to Hilbert-Einstein functional}

\bigskip

In this section we would like to analyze the significance of the
cosmological constant term in the Hilbert-Einstein action for gravity from
the point of view of the G-L model. For this purpose, following Dirac [30]
let us consider the extended Hilbert-Einstein (H-E) action functional $%
S^{c}(g)$ for pure gravity with an extra (cosmological constant) term
defined for some (pseudo) Riemannian manifold $M$ (of total space-time
dimension $d)$ without a boundary: 
\begin{equation}
S^{c}(g)=\int_{M}R\sqrt{g}d^{d}x+C\int_{M}d^{d}x\sqrt{g}\text{ .}  \tag{4.1.}
\end{equation}%
The (cosmological) constant $C$ is determined based on the following chain
of arguments. First, we introduce the following.

\ 

\textbf{Definition 4.1}\textit{.} Let $R_{ij}$\ be the Ricci curvature
tensor. Then the $Einstein$\ $space$\ is defined as a solution of the
following vacuum Einstein equation 
\begin{equation}
R_{ij}=\lambda g_{ij}  \tag{4.2}
\end{equation}%
with $\lambda $\ being constant\textit{.}

\ \ 

From this definition it follows that in both the Riemannian and
pseudo-Riemannian cases 
\begin{equation}
R=d\lambda  \tag{4.3}
\end{equation}%
Following Dirac [30], variation of the action $S^{c}(g)$ produces 
\begin{equation}
R_{ij}-\frac{1}{2}g_{ij}R+\frac{1}{2}Cg_{ij}=0.  \tag{4.4}
\end{equation}%
The combined use of Eq.s(4.3) and (4.4) produces: 
\begin{equation}
C=\lambda (\frac{d-2}{2})  \tag{4.5}
\end{equation}

With these results, using Eq.s(4.3) and (4.5) we can rewrite Eq.(4.4) as
follows: 
\begin{equation}
R_{ij}-\frac{1}{2}g_{ij}R+\frac{1}{2d}(d-2)Rg_{ij}=0.  \tag{4.6}
\end{equation}

\textbf{Remark 4.2}. These observations allow us to look at possible
pseudo-Riemannian extension of the results\ originally obtained by Yamabe,
Refs.[19, 29, 31], for Riemannian manifolds. Although at first sight this
might be of significance importance for high energy physics applications, in
a companion publication [32] we shall argue that dealing with Riemannian
manifolds is already quite sufficient so that the Yamabe results can be used
\ in high energy physics without changes. It should be clear as well that
the earlier introduced notion of the Einstein space is applicable to both
Riemannian and pseudo-Riemannian manifolds.

\ 

In view of this remark, we would like to argue that Eq.(4.6) can also be
obtained by varying the Yamabe functional, Eq.(3.6). Indeed, following Aubin
[29] and Schoen [28], let $t$ be some small parameter labeling the family of
metrics: $g_{ij}(t)=g_{ij}+th_{ij}$. Then, \ these authors demonstrate that 
\begin{equation}
\left( \frac{dR_{t}}{dt}\right) _{t=0}=\bigtriangledown ^{i}\bigtriangledown
^{j}h_{ij}-\bigtriangledown ^{j}\bigtriangledown _{j}h_{i}^{i}-R^{ij}h_{ij} 
\tag{4.7}
\end{equation}%
and 
\begin{equation}
\left( \frac{d}{dt}\sqrt{\left\vert g_{t}\right\vert }\right) _{t=0}=\frac{1%
}{2}\sqrt{\left\vert g\right\vert }g^{ij}h_{ij},  \tag{4.8}
\end{equation}%
where, as usual, $\left\vert g\right\vert $ $=$ $\left\vert \text{det }%
g_{ij}\right\vert .$ Consider now the Yamabe functional, Eq.(3.6), but, this
time, written for the family of metrics which belong to the same conformal
class. We have, 
\begin{equation}
\mathcal{R}(g(t))=\left( V(t)\right) ^{\frac{-2}{p}}\int_{M}R(g(t))DV(t), 
\tag{4.9}
\end{equation}%
where the volume is given by $V(t)=\int_{M}d^{d}x\sqrt{g(t)}$ and,
accordingly, $DV(t)$=$d^{d}x\sqrt{g(t)}.$ Using Eq.s(4.7) and (4.8) in
Eq.(4.9) and taking into account that the combination $\bigtriangledown
^{i}\bigtriangledown ^{j}h_{ij}-\bigtriangledown ^{j}\bigtriangledown
_{j}h_{i}^{i}$ is the total divergence produces the following result: 
\begin{eqnarray}
\left( \frac{d}{dt}\mathcal{R}(g(t))\right) _{t=0} &=&V(0)^{-\frac{2-d}{d}%
}[\int_{M}(Rg^{ij}/2-R^{ij})h_{ij}DV(0)\text{ }\int_{M}DV(0)  \notag \\
&&-(\frac{1}{2}-\frac{1}{d})\int_{M}DV(0)\int_{M}h_{ij}g^{ij}DV(0)]. 
\TCItag{4.10}
\end{eqnarray}%
If the metric $g$ is the critical point of $\mathcal{R}(g(t)),$ then 
\begin{equation}
\left[ R_{ij}-\frac{R}{2}g_{ij}\right] \int_{M}DV(0)+(\frac{1}{2}-\frac{1}{d}%
)(\int_{M}RDV(0))g_{ij}=0.  \tag{4.11}
\end{equation}%
From here, multiplication of both sides by $g^{ij}$ and subsequent summation
produces at once 
\begin{equation}
R-\frac{R}{2}d+(\frac{1}{2}-\frac{1}{d})<R>d=0,  \tag{4.12}
\end{equation}%
where $<R>=\frac{1}{V(0)}\int RDV(0)$ is the average scalar curvature.
Eq.(4.12) can be rewritten as $R=<R>.$ But this condition is \textit{formally%
} equivalent to the Einstein condition, Eq.(4.2), in view of Eq.s(4.3),
(4.6)! Hence, \textit{under such circumstances,} Eq.s (4.11) and (4.12) are
equivalent.

\ 

\textbf{Remark} \textbf{4.3}. It is important to mention that in general the
condition $R=<R>$ is more restrictive than the condition, Eq.(4.2), defining
Einstein spaces. That is to say, the constant Ricci tensor $R_{ij}$ always
leads to the constant scalar curvature $R$ while the Riemannian spaces of
constant scalar curvature $R$ \ are not necessarily spaces for which the
Einstein condition, Eq.(4.2), holds. Hence, they may or may not be of
Einstein type. This was demonstrated by Herglotz in 1916 (immediately after
Einstein's general relativity was completed) [33, page 148]. His results
were subsequently generalized in Refs.[34,35]. We shall \ return to this
topic \ in Section 7 \ where we discuss this problem dynamically \ following
\ the works by Hamilton and Perelman. Using dynamical arguments, we
demonstrate that the equation $R=<R>$ is compatible with the Einstein
condition and, hence describes the Einstein spaces.

In view of this remark, in the rest of this work\ we shall develop our
formalism under the assumption that for the set of metrics of fixed
conformal class, the variational problem for the G-L functional, Eq.(3.7),
is equivalent to the variational problem for the H-E functional, Eq.(4.1),
for pure gravity in the presence of the cosmological constant. Because of
this noted formal equivalence, presence of the cosmological term in the H-E
action amounts to volume conservation for the G-L variational problem. From
here, it should be clear that Eq.(2.12) is equivalent to the condition: $%
R=<R>$ in view of Eq.(3.4)$.$

The result shown in Eq.(4.12) becomes trivial for $d=2$. Physically,
however, the case $d=2$ is important since it is relevant to all known
exactly solvable models of statistical mechanics treatable by methods of
conformal field theories. Hence, now we would like to discuss\ needed
modifications of the obtained results in order to obtain a two dimensional
analog of the G-L theory.

\section{Ginzburg-Landau-like theory in two dimensions}

\subsection{Designing the two dimensional G-L-Yamabe functional}

\ \ From the field-theoretic treatments of the G-L model [36\textbf{]} we
know that \ a straightforward analysis based on asymptotic $\varepsilon -$%
expansions from the critical dimension (4) to the target dimension (2) is
impractical. At the same time, the results of CFT and exactly solvable
models make sense thus far only in $d=2$. The question arises: is there an
analog of G-L Eq.(3.1) in two dimensions? And, if such an analog does exist,
what use can be made of such a result? In this section we provide
affirmative answers to these questions. We demonstrate that: a) indeed, such
a two dimensional analog of the G-L equation does exist and is given by the
Liouville Eq.(5.19), b) the functional, Eq.(5.14), whose minimization
produces such an equation is the exact two dimensional analog of the
G-L-Yamabe functional, Eq.(3.2), c) these results can be (re)obtained from
the existing string-theoretic formulations of the CFT\ which were developed
entirely independently.

\QTP{Body Math}
To discuss topics related to items a) and b) just mentioned, we begin with
the observation that in two dimensions, Eq.(2.4) acquires a very simple form 
\begin{equation}
\Delta _{\hat{g}}=e^{-2f}\Delta _{g}\text{ ,}  \tag{5.1}
\end{equation}%
where we use a factor of 2 to be in accord with Eq.(2.11) for the scalar
curvature. According to Eq.(2.11), the scalar curvature in two dimensions
transforms like 
\begin{equation}
\hat{R}(\hat{g})=e^{-2f}\{R(g)-\Delta _{g}2f\}  \tag{5.2}
\end{equation}%
while the area $dA=d^{2}x\sqrt{g}$ transforms like 
\begin{equation}
d\hat{A}=e^{-2f}dA.  \tag{5.3}
\end{equation}%
These facts immediately suggest that the previously introduced action
functional 
\begin{equation}
S[\mathbf{X}]=\int_{M}d^{2}x\sqrt{g}(\bigtriangledown _{g}\mathbf{X})\cdot
(\bigtriangledown _{g}\mathbf{X})  \tag{5.4}
\end{equation}%
is conformally invariant. Using results by Polyakov [37] and noting that $%
(\bigtriangledown _{g}\mathbf{X})\cdot (\bigtriangledown _{g}\mathbf{X}%
)=g^{\alpha \beta }\partial _{\alpha }X^{\mu }\partial _{\beta }X_{\mu }$,
we need to consider the following path integral\footnote{%
Without loss of generality, we would like to consider the case of one
component field $\phi $ only.} 
\begin{equation}
\exp (-\mathcal{F(}\mathit{g}))\equiv \int D[\phi ]\exp (-\frac{1}{2}%
\int_{M}d^{2}x\sqrt{g}g^{\alpha \beta }\partial _{\alpha }\phi \partial
_{\beta }\phi ).  \tag{5.5}
\end{equation}%
Here the symbol $\mathcal{F}$($g$) stands for the \textquotedblright free
energy\textquotedblright\ usually defined this way in statistical mechanics%
\footnote{%
Usually, instead of $\mathcal{F(}\mathit{g)}$ one writes $\mathcal{F(}%
\mathit{g)/k}_{B}\mathit{T,}$where $T$ is the temperature and $k_{B}$ is the
Bolzmann's constant. In the present case the problems we are studying do not
require specific values for these constants. For this reason they will be
suppressed.}. Fortunately, this integral was calculated by Polyakov [37%
\textbf{]} for two dimensional manifolds $M$ \textit{without} boundaries and
by Alvarez [38] for manifolds \textit{with} boundaries. In this work, to
avoid unnecessary complications, we shall be concerned with manifolds 
\textit{without} boundaries. Although in the original work by Polyakov, one
can find the final result of calculation of the above path integral, the
details of this calculation can only be found elsewhere. In particular, we
shall follow pedagogically written papers by Weisberger [39,40\textbf{]} and
Osgood, Phillips and Sarnak [41-43] (OPS).

\ To begin, let $\hat{g}_{\alpha \beta }$ be some reference metric, while $%
g_{\alpha \beta }$ is a metric conformally related to it, i.e. $g_{\alpha
\beta }=\exp (-2\varphi )\hat{g}_{\alpha \beta }$ . Should the above path
integral be for the flat (i.e.$g_{\alpha \beta }$=$\delta _{\alpha \beta }$
) two dimensional manifold, one would have at once the result: $\mathcal{F}$=%
$\frac{1}{2}\ln \det^{\prime }\Delta _{0}$,\ where the prime indicates that
the zero mode is omitted. Because the\ flatness assumption in general is
physically incorrect, it is appropriate to pose a problem: how is the path
integral for the metric $g$ related to the path integral for the metric $%
\hat{g}$ ? The paper [41] by OPS provides an answer, e.g. see Eq.(1.13) of
this reference. To connect this equation with that known in the physics
literature we replace it by the equivalent expression 
\begin{equation}
\ln \left( \frac{\det^{\prime }\Delta _{\hat{g}}}{A_{\hat{g}}}\right) -\ln
\left( \frac{\det^{\prime }\Delta _{g}}{A_{g}}\right) =-\frac{1}{6\pi }[%
\frac{1}{2}\int\limits_{M}dA_{g}\{\left\vert \bigtriangledown _{g}\varphi
\right\vert ^{2}+R(g)\varphi \}]  \tag{5.6}
\end{equation}%
useful in applications to strings and CFT, e.g see Ref.[13], page 637%
\footnote{%
In arriving at this result we took into account that the Gaussian curvature $%
K$ in Ref.[41] is related to Riemannian curvature $R$ as $R=2K$.}.

It is worthwhile to provide a few computational details leading to Eq.(5.6).
Let, therefore, $\hat{\psi}_{i}$ be eigenfunctions of the Laplacian $\Delta
_{g}$ with eigenvalues $\lambda _{i}$ arranged in such a way that $0=\hat{%
\lambda}_{0}<\hat{\lambda}_{1}\leq \hat{\lambda}_{2}\leq \cdot \cdot \cdot ,$
i.e. 
\begin{equation}
-\Delta _{g}\hat{\psi}_{i}+\hat{\lambda}\hat{\psi}_{i}=0.  \tag{5.7}
\end{equation}%
Then, we construct the zeta function 
\begin{equation}
\zeta _{g}(s)=\sum\limits_{i=1}^{\infty }\hat{\lambda}_{i}^{-s}  \tag{5.8}
\end{equation}%
in such a way that 
\begin{equation}
\det \text{ }^{\prime }\Delta _{g}=\exp (-\zeta _{g}^{^{\prime }}(0)) 
\tag{5.9}
\end{equation}%
with $\zeta _{g}^{^{\prime }}(0)=\left( \frac{d}{ds}\zeta _{g}(s)\right)
_{s=0}.$ Using Eq.(5.1), we obtain as well 
\begin{equation}
-e^{-2\varphi }\Delta _{g}\psi _{i}+\lambda \psi _{i}=0.  \tag{5.10}
\end{equation}%
In particular, for a constant $\varphi =\bar{\varphi}$ we obtain 
\begin{equation}
\zeta _{\hat{g}}(s)=\sum\limits_{i=1}^{\infty }\left( e^{-2\bar{\varphi}}%
\hat{\lambda}_{i}\right) ^{-s}=e^{2s\bar{\varphi}}\zeta _{g}(s).  \tag{5.11}
\end{equation}%
Use of Eq.(5.9) in Eq.(5.11) produces: 
\begin{equation}
\zeta _{\hat{g}}^{^{\prime }}(0)=\zeta _{g}^{^{\prime }}(0)+2\bar{\varphi}(%
\frac{\chi (M)}{6}-1).  \tag{5.12}
\end{equation}%
This result was obtained with help of the known fact, Ref.[41], Eq.(1.9),
that 
\begin{equation}
\zeta _{g}(0)=\frac{\chi (M)}{6}-1  \tag{5.13}
\end{equation}%
with $\chi (M)$ being Euler's characteristic of a two dimensional manifold $%
M $ without boundaries. In view of the definition, Eq.(5.9), and the
Gauus-Bonnet theorem,\ we observe that Eq.(5.6) is reduced to Eq.(5.12) for
the case of constant conformal factor $\varphi =\bar{\varphi}$, as required.
\ 

\QTP{Body Math}
Using Eq.(5.6) and, following OPS [41], we would like to consider the
related functional $\mathcal{F}$($\varphi )$ defined by 
\begin{equation}
\mathcal{F}(\varphi )=\frac{1}{2}\int\limits_{M}dA_{g}\{\left\vert
\bigtriangledown _{g}\varphi \right\vert ^{2}+R(g)\varphi \}-\pi \chi (M)\ln
\int\limits_{M}dA_{g}\text{ }e^{2\varphi }.  \tag{5.14}
\end{equation}%
This functional is the exact analog of the Yamabe functional, Eq.(3.2), in
dimensions 3 and higher as we shall demonstrate below and in Section\textbf{%
\ }7. In the meantime, in view of Eq.(5.6), it can be rewritten in the
following equivalent form\footnote{%
Here one should understand the word "equivalence" in the sence that both
functionals produce the same critical metrics upon minimization.} 
\begin{equation}
\mathcal{F}(\varphi )=-6\pi \ln \det \text{ }^{\prime }\Delta _{g}+\pi
(6-\chi (M))\ln A.  \tag{5.15}
\end{equation}%
Let $a$ be some constant, then 
\begin{equation}
\mathcal{F}(\varphi +a)=\mathcal{F}(\varphi ).  \tag{5.16}
\end{equation}%
Such invariance signifies the fact that the above action is scale invariant.
This property is in complete accord with Eq.s(2.7) and (2.8). If we impose a
constraint (fix the gauge) : $A=1$, then we would end up with the
Liouville-like action used in CFT. We would like to explain all this in some
detail now.

Following OPS, it is convenient to replace the constraint $A=1$ by the
alternative constraint on the field $\varphi $: 
\begin{equation}
\int\limits_{M}\varphi dA_{g}=0.  \tag{5.17}
\end{equation}%
To demonstrate that such a constraint is equivalent to the requirement $A=1$%
, we note that, provided that the field $\psi $ minimizes $\mathcal{F}(\psi
) $ subject to the constraint Eq.(5.17), the field 
\begin{equation}
\varphi =\psi -\frac{1}{2}\ln \int\limits_{M}\exp (2\psi )dA_{g}  \tag{5.18}
\end{equation}%
minimizes $\mathcal{F}(\varphi )$ subject to the constraint $A=1$. Using
these facts, minimization of $\mathcal{F}(\psi )$ produces the following
Liouville equation 
\begin{equation}
-\Delta _{g}\psi +\frac{1}{2}R(g)-\frac{2\pi \chi (M)\exp (2\psi )}{%
\int\limits_{M}\exp (2\psi )dA_{g}}=0.  \tag{5.19}
\end{equation}%
Comparing this result with Eq.(5.2) and taking into account that 
\begin{equation}
\frac{2\pi \chi (M)}{\int\limits_{M}\exp (2\psi )dA_{g}}=\hat{R}(\hat{g}%
)=const  \tag{5.20}
\end{equation}%
we conclude that, provided that the background metric $g$ is given so that
the scalar curvature $R(g)$ (not necessarily constant) can be calculated,
the Liouville Eq.(5.19) is exactly analogous to that previously obtained in
Eq.(3.3) (which is the same as Eq.(2.12)). In view of this analogy, use of
Eq.s.(3.4) and (3.6) as well as Eq.s(5.14),(5.15) causes the functional $%
\mathcal{F}(\varphi )$ to attain its extremum for metric $\hat{g}$ of
constant scalar curvature $\hat{R}(\hat{g})$. To decide if the extremum is
minimum or maximum we have to consider separately the cases $\chi (M)>0$ and 
$\chi (M)\leq 0.$ In the case of $\chi (M)>0$ we have only to consider
manifolds homeomorphic to $S^{2}$ so that $\chi (M)=2$. Fortunately, this
case was considered in detail by Onofri [44]\footnote{%
Generalization of these two dimensional results to higher dimensional
manifolds of even dimensions can be found in the paper by Beckner [45].}.
Using his work, the following inequality 
\begin{equation}
\ln \int\limits_{S^{2}}dA_{\hat{g}}\exp (\psi )\leq \int\limits_{S^{2}}dA_{%
\hat{g}}\psi +\frac{1}{4}\int\limits_{S^{2}}dA_{\hat{g}}\left\vert
\bigtriangledown \psi \right\vert ^{2}  \tag{5.21}
\end{equation}%
attributed to Aubin [29] and inspired by earlier results by Trudinger and
Moser [45], is helpful for deciding whether the obtained extremum is a
minimum or maximum. Here $\hat{g}$ is the metric of the unit sphere $S^{2}$
with a constant Gaussian curvature of $1$. The metric $g$ conformal to $\hat{%
g}$ is given by $g=\exp (2\psi )\hat{g}$ , with $\psi $ obeying the
Liouville equation (just like Eq.(5.19)), where both $R(g)$ and $R(\hat{g})$
are constant by virtue of the initial choice of $\hat{g}$. By combining
Eq.s(5.6),(5.14) and (5.15) with the inequality (5.21) \ and taking into
account that \ by design ln$A_{\hat{g}}=0$ we obtain, 
\begin{equation}
-3\pi \ln \frac{\det^{\prime }\Delta _{g}}{\det^{\prime }\Delta _{\hat{g}}}=%
\frac{1}{4}\int\limits_{S^{2}}dA_{\hat{g}}\{\left\vert \bigtriangledown _{%
\hat{g}}\psi \right\vert ^{2}+2\psi \}-\ln \int\limits_{S^{2}}dA_{\hat{g}%
}\exp (2\psi )\geq 0,  \tag{5.22}
\end{equation}%
with equality occurring only at the extremum $\psi =\psi ^{\ast },$ with
function $\psi ^{\ast }$ being a solution of the Liouville Eq.(5.19). It can
be shown [44] that: a) such a solution involves only M\"{o}bius
transformations of the sphere $S^{2}$ and that, b) the functional $\mathcal{F%
}(\varphi )$ is invariant with respect to such transformations. The case of $%
\chi (M)\leq 0$ \ is treated in Section 2.2 of the OPS paper, Ref.[41] and
leads to the same conclusions about the extremality of the functional $%
\mathcal{F}(\varphi ).$

\ 

\textbf{Corrollary 5.1.} \ The two dimensional results just obtained \ 
\textsl{by design} are in accord with results obtained in higher dimensions
( discussed in Sections 3,6 and to be discussed in Section 7). In particular,%
\textsl{\ the functional} $\mathcal{F}(\varphi )$ \textsl{is the exact
analog of the Yamabe functional} $S$[$\varphi ],$ Eq.(3.2). Since in both
cases the functionals are \textquotedblright
translationally\textquotedblright\ (actually, \textquotedblright
scale\textquotedblright ) invariant, e.g. compare Eq.(3.6) with Eq.(5.16),
in both cases, the extremum is realized for the metric conformal to the
metric of constant scalar curvature, e.g. compare Eq.(3.3) with the
Liouville Eq.(5.19).

\subsection{\ \ Connections with string and CFT}

\bigskip\ 

The obtained results allow us now to discuss topic c) listed at the
beginning of this section. Using known results for string and conformal
field theories, the results just obtained can be given a statistical
mechanical interpretation. Following [14,15,46], it is of interest to
consider averages of the vertex operators. They are defined by 
\begin{equation}
\left\langle \prod\limits_{i=1}^{n}\exp (\beta _{i}\phi
(z_{i}))\right\rangle \equiv \int D\left[ \phi \right] \exp \{-S_{L}(\phi
)\}\prod\limits_{i=1}^{n}\exp (\beta _{i}\phi (z_{i})),  \tag{5.23}
\end{equation}%
where the Liouville action $S_{L}(\phi )$ is given (in notations adopted
from these references) by 
\begin{equation}
S_{L}(\phi )=\frac{1}{8\pi }\int_{M}dA_{\hat{g}}[\left\vert \bigtriangledown
\phi \right\vert ^{2}-QR(\hat{g})\phi +8\pi \bar{\mu}\exp (\alpha _{+}\phi
)].  \tag{5.24}
\end{equation}%
The actual values and the meaning of the constants $Q$, $\bar{\mu}$ and $%
\alpha _{+}$ are explained in these references and are of no immediate use
for us. Clearly, upon proper rescaling, we can bring $S_{L}(\phi )$ to the
form which agrees with $\mathcal{F}(\varphi ),$ defined by Eq.(5.14),
especially in the trivial case when both $\chi (M)$ and $\bar{\mu}$ are
zero. When they are not zero, the situation in the present case becomes
totally analogous to that discussed earlier for the Yamabe functonal. In
particular, in Section 3 we noted that the G-L Euler-Lagrange Eq.(3.3) can
be obtained either by minimization of the Yamabe functional, Eq.(3.2), (or
(4.9)) or by minimization of the G-L functional, Eq.(3.7), where the
coupling constant $\lambda $ plays the role of the Lagrange multiplier
keeping track of the volume constraint. In the present case, variation of
the Liouville action $S_{L}(\phi )$ will produce the Liouville equation,
e.g. see Eq.s(5.19)-(5.20), which is the two dimensional analog of the G-L
Eq.(3.3). Such variation is premature, however, since we can reobtain $%
\mathcal{F}(\varphi )$ exactly \ using the path integral, Eq.(5.23). This
procedure then will lead us directly to the Liouville Eq.(5.19).

For this purpose, we need to consider the path integral, Eq.(5.23), in the
absence of sources, i.e. when all $\beta _{i}=0.$ Following ideas of Ref.s
[14,15,46], we take into account that: a) 
\begin{equation}
\frac{1}{4\pi }\int\limits_{M}dA_{\hat{g}}R(\hat{g})=\chi (M)=2-2g 
\tag{5.26}
\end{equation}%
with $g$ being genus of $M$ and, b) the field $\phi $ can be decomposed into 
$\phi =\phi _{0}+\varphi $ in such a way that $\phi _{0\text{ }}$is
coordinate-independent and $\varphi $ is subject to the constraint given by
Eq.(5.17). Then, use of the identity 
\begin{equation}
\int\limits_{-\infty }^{\infty }dx\exp (ax)\exp (-b\exp (\gamma x))=\frac{1}{%
\gamma }b^{-\frac{a}{\gamma }}\Gamma (\frac{a}{\gamma })  \tag{5.27}
\end{equation}%
(with $\Gamma (x)$ being Euler's gamma function) in the path integral,
Eq.(5.23), requires us to evaluate the following integral 
\begin{eqnarray}
I &=&\int\limits_{-\infty }^{\infty }d\phi _{0}\exp (\phi _{0}\frac{Q}{2}%
\chi (M))\exp ((\bar{\mu}\int\limits_{M}dA_{\hat{g}}\exp (\alpha _{+}\varphi
))\exp (\alpha _{+}\phi _{0})  \notag \\
&=&\frac{\Gamma (-s)}{\alpha _{+}}(\bar{\mu}\int\limits_{M}dA_{\hat{g}}\exp
(\alpha _{+}\varphi ))^{s}  \TCItag{5.28}
\end{eqnarray}%
with $s$ being given by 
\begin{equation}
s=-\frac{Q}{2\alpha _{+}}\chi (M).  \tag{5.29}
\end{equation}%
Using this result in Eq.(5.23), we obtain the following path integral (up to
a constant) 
\begin{equation}
Z[\varphi ]=\int D[\varphi ]\exp (-\mathcal{\hat{F}}(\varphi ))  \tag{5.30}
\end{equation}%
with functional $\mathcal{\hat{F}}(\varphi )$ given by 
\begin{equation}
\mathcal{\hat{F}}(\varphi )=S_{L}(\varphi ;\bar{\mu}=0)-\frac{Q}{2\alpha _{+}%
}\chi (M)\ln [\bar{\mu}\int\limits_{M}dA_{\hat{g}}\exp (\alpha _{+}\varphi
)].  \tag{5.31}
\end{equation}%
This functional (up to rescaling of the field $\varphi $) is just the same
as the Yamabe-like functional $\mathcal{F}(\varphi )$ given by Eq.(5.14).

Define now the free energy $\mathcal{F}$ in the usual way via $\mathcal{F}%
=-\ln Z[\varphi ]$ (as was done after Eq.(5.5)) and consider the saddle
point approximation to the functional integral $Z[\varphi ].$ Then, \ for
spherical topology in view of Eq.(5.22), we (re)obtain $\mathcal{F}\geq 0$
with equality obtained exactly when $\varphi =\psi ^{\ast }$. Inclusion of
the sources (or the vertex operators) can be taken into account also,
especially in view of the results of [43].

To understand better the physical significance of the obtained results, it
is useful to reobtain them using a somewhat different method. The results
obtained with this alternative method are also helpful when we shall discuss
their higher dimensional analogs in the next section. For this purpose we
write (up to a normalization constant) 
\begin{equation}
\int D\left[ \phi \right] \exp \{-S_{L}(\phi )\}=\int\limits_{0}^{\infty
}dAe^{-\bar{\mu}A}Z_{L}(A)  \tag{5.32}
\end{equation}%
where 
\begin{equation}
Z_{L}(A)=\int D[\phi ]\delta (\int_{M}dA_{\hat{g}}\exp (\alpha _{+}\phi
)-A)\exp (S_{L}(\phi ;\bar{\mu}=0)).  \tag{5.33}
\end{equation}%
If, as before, we assume that $\phi =\phi _{0}+\varphi ,$ then an elementary
integration over $\phi _{0}$ produces the following explicit result for $%
Z_{L}(A):$%
\begin{equation}
Z_{L}(A)=\frac{-1}{\alpha _{+}}A^{\omega }\int D[\varphi ]\left[ \int_{M}dA_{%
\hat{g}}\exp (\alpha _{+}\varphi )\right] ^{-(\omega +1)}\exp
(-S_{L}(\varphi ;\bar{\mu}=0))  \tag{5.34}
\end{equation}%
where the exponent $\omega $ is given by 
\begin{equation}
\omega =\frac{\chi (M)Q}{2\alpha _{+}}-1.  \tag{5.35}
\end{equation}%
Finally, using Eq.(5.34) in Eq.(5.32) produces back Eq.s(5.30) and (5.31)
(again, up to a constant factor). An overall \textquotedblright
-\textquotedblright\ sign can be removed by proper normalization of the path
integral. \ These results can be used for computation of the correlation
functions of conformal field theories (CFT). Details can be found in
[14,15,46].

\bigskip

\section{\protect\bigskip Designing higher dimensional CFT(s)}

\subsection{General remarks}

\textit{\ }In previous section we\textit{\ }explained a delicate
interrelationship between the path integrals Eq.(5.5) and (5.23). From the
literature on CFT cited earlier it is known that, actually, \textit{both}
are being used for the design of different CFT\ models in 2 dimensions. For
instance, if one entirely ignores the effects of curvature in Eq.(5.5), then
one ends up with the Gaussian -type path integral whose calculation for a
flat torus is discussed in detail in Ref.[21], pages 340-343, and, by
different methods, in our Ref.[47]. By making appropriate changes of the
boundary conditions in such a path integral (or, equivalently, by
considering appropriately chosen linear combinations of modular invariants)
it is possible to build partition functions for all existing CFT models. For
the same purpose one can use the path integral given by Eq.(5.23) but the
calculation proceeds differently as explained in \ Section 5. Since in 2
dimensions the conformal invariance is crucial in obtaining exact results,
use of the Gaussian-type path integrals is, strictly speaking, not
permissible. Fortunately, the saddle point -type calculations made for the
path integral, Eq.(5.23), produce the same results since the extremal
metrics happen to be flat [47]. If one does not neglect curvature effects in
Eq.(5.5), one eventually ends up with the integrand of the path integral,
Eq.(5.23). This result is a consequence of Eq.(5.6) known as the conformal
anomaly. If one would like to proceed \textit{in an analogous fashion} in
dimensions higher than two one should be aware that there is a profound
difference between calculations done in odd and even dimensions.We would
like to explain this circumstance in some detail now.

In 2 dimensions the conformal invariance of the action, Eq.(5.4), has been
assured by the transformational properties of the 2 dimensional Laplacian
given by Eq.(5.1). In higher dimensions, the Laplacian is transformed
according to Eq.(2.4), so that even the simplest Gaussian model is not
conformally invariant! This observation makes use of traditional string
-theoretic methods in higher dimensions problematic. In two dimensions these
are based on a two stage process: first one calculates the path integral,
Eq.(5.5), exactly and, second, one uses the result of such a calculation
(the conformal anomaly) as an input in another path integral, e.g.
Eq.(5.23), which is obtained by integrating this input over all members of
the conformal class. Since in odd dimensions there is no conformal anomaly
as we shall demonstrate momentarily, such a two stage process cannot be
used. The situation can be improved considerably if we \textit{do not} rely
on the use of the two stage process just described. We would like to explain
this fact in some detail now.

Even though the transformational properties (with respect to conformal
transformations) of the Laplacian, Eq.(2.4), in dimensions higher than 2 are
rather unpleasant, fortunately, they can be considerably improved if,
instead of the usual Laplacian, one uses the conformal (Yamabe) Laplacian $%
\square _{g}$ defined by 
\begin{equation}
\square _{g}=\Delta _{g}+\hat{\alpha}R(g),  \tag{6.1}
\end{equation}%
where $\hat{\alpha}=\alpha ^{-1}=\frac{1}{4}\frac{d-2}{d-1}.$ By
construction, in 2 dimensions it becomes the usual Laplacian. In higher
dimensions its transformational properties are much simpler than those for
the usual Laplacian (e.g see Eq.(2.4)). Indeed, it can be shown [48] that 
\begin{equation}
\square _{e^{2f}g}=e^{-(\frac{d}{2}+1)f}\square _{g}e^{(\frac{d}{2}-1)f}. 
\tag{6.2}
\end{equation}%
This result can be easily understood if we use the results of Section 2.
Indeed, since $e^{2f}=\varphi ^{p-2}$ and since $p=\frac{2d}{d-2}$, we
obtain at once $e^{(\frac{d}{2}-1)f}=\varphi $, while the factor $e^{-(\frac{%
d}{2}+1)f}$ is transformed into $\varphi ^{1-p}.$ From here, it is clear
that Eq.(2.12) for scalar curvature can be equivalently rewritten as 
\begin{equation}
\hat{R}(\hat{g})=\alpha \varphi ^{1-p}(\Delta _{g}\varphi +\alpha
^{-1}R(g)\varphi ),  \tag{6.3}
\end{equation}%
where $\hat{g}=e^{2f}g$ , so that 
\begin{equation}
\hat{R}(\hat{g})=\alpha \square _{e^{2f}g}\text{ .}  \tag{6.4}
\end{equation}%
In practical applications it could be more useful to consider two successive
conformal transformations made with the conformal factors $e^{2f}$ and $%
e^{2h}.$ If $e^{2f}=\varphi ^{p-2}$ and $e^{2h}=\psi ^{p-2}$ then, we
obtain, 
\begin{equation}
\Delta _{\hat{g}}\psi +\alpha ^{-1}R(\hat{g})\psi =\varphi ^{1-p}[\Delta
_{g}\left( \varphi \psi \right) +\alpha ^{-1}R(g)\left( \varphi \psi \right)
].  \tag{6.2b}
\end{equation}%
This result is, of course, equivalent to Eq.(6.2a).

Consider now the following path integral 
\begin{equation}
\exp \left( -\mathcal{F}(g)\right) =\int D\left[ \varphi \right] \exp
\{-S_{\Box _{g}}(\varphi )\}  \tag{6.5}
\end{equation}%
where 
\begin{eqnarray}
S_{\Box _{g}}(\varphi ) &=&\alpha \int_{M}d^{d}x\sqrt{g}\{(\bigtriangledown
_{g}\varphi )^{2}+\alpha ^{-1}R(g)\varphi ^{2}\}  \notag \\
&=&\alpha \int_{M}d^{d}x\sqrt{g}[\varphi \square _{g}\varphi ]=\int_{M}d^{d}x%
\sqrt{\tilde{g}}\hat{R}(\hat{g})  \TCItag{6.6}
\end{eqnarray}%
The conformal factor $\varphi ^{-p}$ in Eq.(6.3) is eliminated by the
corresponding factor coming from the volume factor of $\sqrt{\tilde{g}}=%
\sqrt{\exp (2f)g}=\varphi ^{p}\sqrt{g}\footnote{%
This result should be understood as follows. We have $\tilde{g}_{ij}=\varphi
^{p-2}g_{ij}.$From here, we have for determinants$\sqrt{\tilde{g}}=[\varphi
^{p-2}]^{\frac{d}{2}}\sqrt{g}=\varphi ^{p}\sqrt{g}$ in accord with Eq.(3.5).}%
.$

\textbf{Corollary 6.1}. Thus, Eq.(6.5)\textit{\ is the exact higher
dimensional analog of the two dimensional path integral}, Eq.(5.5), and,
hence, problems related to higher dimensional CFT are those of Riemannian%
\footnote{%
Sometimes called Euclidean quantum gravity.} quantum gravity [49]. In this
paper we are not considering the pseudo-Riemannian case relevant for true
Einsteinian gravity. It will be discussed in a companion publication [32].

The question immediately arises: If the path integral, Eq.(6.5), is such an
analog, is there a higher dimensional analog of Eq.(5.6)? The answer is
\textquotedblright yes\textquotedblright , if the dimension of space is
even, and \textquotedblright no\textquotedblright\ if the dimension of space
is odd [48].

\ 

\subsection{\ Lack of conformal anomaly in odd dimensions}

\ 

In view of its importance, we would like to provide a sketch of the
arguments leading to the answer \textquotedblright no\textquotedblright\ \
in dimension 3. Clearly, the same kind of arguments \ will apply in any
other odd dimension. In doing so, although we follow the arguments of Refs.
[48,50], some of our derivations are original. We begin by assuming that
there is a one parameter family of metrics: $\hat{g}(x)$ =exp$(2xf)g$. Next,
we define the operator $\delta _{f}$ via 
\begin{equation}
\delta _{f}\square _{g}=\frac{d}{dx}\mid _{x=0}\square _{\text{exp}(2xf)g}%
\text{ .}  \tag{6.7}
\end{equation}%
Taking into account Eq.(6.2) we obtain explicitly 
\begin{equation}
\delta _{f}\square _{g}=-2f\square _{g}  \tag{6.8}
\end{equation}%
and, accordingly, 
\begin{equation}
\delta _{f}e^{-t\square _{g}}=-t(\delta _{f}\square _{g})e^{-t\square _{g}}.
\tag{6.9}
\end{equation}%
These results allow us to write for the zeta function\footnote{%
E.g. see Eq.(5.8) and take into account the identity $x^{-s}\Gamma
(s,x)=\int\limits_{0}^{\infty }dyy^{s-1}\exp (-xy).$} 
\begin{eqnarray}
\delta _{f}\zeta _{\square _{g}}(s) &=&\frac{1}{\Gamma (s)}%
\int\limits_{0}^{\infty }dtt^{s-1}\delta _{f}Tr(e^{-t\square _{g}})  \notag
\\
&=&\frac{1}{\Gamma (s)}\int\limits_{0}^{\infty }dtt^{s}Tr(-2f\square
_{g}e^{-t\square _{g}})  \notag \\
&=&\frac{-2s}{\Gamma (s)}\int\limits_{0}^{\infty }dtt^{s-1}Tr(fe^{-t\square
_{g}}).  \TCItag{6.10}
\end{eqnarray}%
The last line was obtained by performing integration by parts. Since for
small $t^{\prime }$s it is known that, provided that $\dim \ker \square
_{g}=0\footnote{%
In the case of path integrals this is always assumed since zero modes of the
corresponding operators are associated with some kind of translational,
rotational, etc. symmetry. To eliminate the undesirable dilatational
symmetry, one actually should use the Yamabe functional, Eq.(3.6), as
explained in Section 3. Although this is silently assumed thus far,
arguments additional to those in Section 3 will be introduced further below.}%
,$ 
\begin{equation}
Tr(fe^{-t\square _{g}})\simeq \sum\limits_{k=0}^{\infty }\left(
\int\limits_{M}f(x)u_{k}(x)dvol\right) t^{k-d/2},  \tag{6.11}
\end{equation}%
where $dvol=\sqrt{g}d^{d}x,$ as usual. Using this result in Eq.(6.10)
produces 
\begin{eqnarray}
\delta _{f}\zeta _{\square _{g}}(0) &=&-\frac{2s}{\Gamma (s)}%
(\int\limits_{0}^{1}dtt^{s-1}Tr(fe^{-t\square _{g}})+\int\limits_{1}^{\infty
}dtt^{s-1}Tr(fe^{-t\square _{g}}))  \notag \\
&=&-\frac{2s}{\Gamma (s)}(\sum\limits_{k=0}^{\infty }\frac{%
\int\nolimits_{M}fu_{k}dvol}{s+k-d/2}+\int\limits_{1}^{\infty
}dtt^{s-1}Tr(fe^{-t\square _{g}}))\mid _{s=0}.  \TCItag{6.12}
\end{eqnarray}%
Since for $s\rightarrow 0^{+}\ $we have$\ \ \left( 1/\Gamma (s)\right) \sim
s,$ the second (regular) term in brackets will become zero when multiplied
by the combination $\frac{2s}{\Gamma (s)}$ , while the first term will
become zero even if it might acquire a pole (in case if $d=2k$). Hence, for
all dimensions $d\geq 3$ we obtain $\delta _{f}\zeta _{\square _{g}}(0)=0$
or 
\begin{equation}
\zeta _{\square _{g}}(0)=\zeta _{\square _{\hat{g}}}(0).  \tag{6.13}
\end{equation}%
This result can be used further. Indeed, if we write 
\begin{equation}
\delta _{f}\left[ \Gamma (s)\zeta _{\square _{g}}(s)\right] =\Gamma
(s)[\delta _{f}\zeta _{\square _{g}}(0)+s\delta _{f}\zeta _{\square
_{g}}^{\prime }(0)+O(s^{2})]  \tag{6.14}
\end{equation}%
and take into account Eq.(6.13) and the fact that $s\Gamma (s)=1$ we obtain, 
\begin{eqnarray}
\delta _{f}\zeta _{\square _{g}}^{\prime }(0) &=&\delta
_{f}\int\limits_{0}^{\infty }dtt^{s-1}Tr(e^{-t\square _{g}})\mid _{s=0} 
\notag \\
&=&-2s(\sum\limits_{k=0}^{\infty }\frac{\int\nolimits_{M}fu_{k}dvol}{s+k-d/2}%
+\int\limits_{1}^{\infty }dtt^{s-1}Tr(fe^{-t\square _{g}}))\mid _{s=0}. 
\TCItag{6.15}
\end{eqnarray}%
Applying the same arguments to Eq.(6.15) as those that were used for
Eq.(6.12) we conclude that, provided $\dim \ker \square _{g}=0,$ in \textit{%
odd} dimensions, $\delta _{f}\zeta _{\square _{g}}^{\prime }(0)=0,$ that is%
\begin{equation}
\zeta _{\square _{g}}^{\prime }(0)=\zeta _{\square _{\hat{g}}}^{\prime }(0).
\tag{6.16}
\end{equation}%
In view of the Eq.(5.9) this leads also to 
\begin{equation}
\det \square _{g}=\det \square _{\hat{g}},  \tag{6.17}
\end{equation}%
QED.

\subsection{3d CFT\ path integrals}

\subsubsection{General remarks}

Our previously obtained results can be further refined if we recall
Eq.s(5.9)-(5.11). In particular, let $e^{2\bar{\varphi}}$ in Eq.(5.11) be
rewritten as some nonnegative constant $l$ . Then we obtain 
\begin{equation}
\zeta _{\hat{g}}(s)=l^{s}\zeta _{g}(s).  \tag{6.18}
\end{equation}%
This result is consistent with Eq.(6.13) for $s=0.$ Differentiation with
respect to $s$ produces 
\begin{equation}
\zeta _{\hat{g}}^{\prime }(0)=\zeta _{g}(0)\ln l+\zeta _{g}^{\prime }(0). 
\tag{6.19}
\end{equation}%
In view of Eq.(5.9), this result is equivalent to lndet $\square _{\hat{g}%
}=\ln $det $\square _{g}-\zeta _{g}(0)\ln l$. and apparently contradicts
Eq.(6.17). But the contradiction is only apparent in view of the earlier
footnote. The situation is easily correctable if in the path integral,
Eq.(6.5), we replace the action functional $S_{\Box }(\varphi )$ by that of
Yamabe given by Eq.s(3.2) (or (3.6)). This, by the way, allows us also to
fix the value of $\zeta _{g}(0)$: provided that we identify the constant $l$
with the volume $V$, the value of $\zeta _{g}(0)=\frac{2}{p}.$ After this,
the situation\ in the present case becomes very similar to that encountered
in the previous section. That is, instead of the functional $\mathcal{F}%
(\varphi )$ given by Eq.(5.15), we can consider now the related functional
given by 
\begin{eqnarray}
\mathcal{F}(\varphi ) &=&\ln \text{det}\square _{g}-\zeta _{g}(0)\ln V 
\notag \\
&\equiv &\ln \text{det}\square _{g}-\frac{2}{p}\ln V.  \TCItag{6.20}
\end{eqnarray}%
For the path integral calculations, a functional defined in such a way is
not yet sufficient. To repair this deficiency we have to impose a volume
constraint. That is we need to consider the path integral of the type 
\begin{equation}
Z_{Y}(V)=\int D[\varphi ]\delta (\int_{M}d^{d}x\sqrt{g}\varphi ^{p}-V)\exp
(-S[\varphi ])  \tag{6.21}
\end{equation}%
with $S[\varphi ]$ given by Eq.(3.2) (or (3.6)).

\ 

\textbf{Remark 6.2.} The path integral $Z_{Y}(V)$ ($Y$ in honor of Yamabe)
is the exact higher dimensional analog of the \textquotedblright
stringy\textquotedblright\ path integral $S_{L}(A)$ given by Eq.(5.33). In
view of Eq.(3.6), it also can be viewed as the path integral for pure
(Euclidean) gravity in the presence of the cosmological constant\footnote{%
This fact can be used (with some caution in view of Remark 4.3.) as
alternative formulation of the quantum gravity problem.}.

\ \ 

In complete analogy with Eq.(5.32), the standard path integral for the self
interacting scalar $\varphi ^{4}$ (or LGW) field theory is obtainable now as
follows 
\begin{equation}
\int D\left[ \phi \right] \exp \{-S_{LGW}(\phi )\}=\int\limits_{0}^{\infty
}dVe^{-bV}Z_{Y}(V).  \tag{6.22}
\end{equation}%
But, since the variation of the Yamabe functional produces the same Eq.(3.3)
as can be obtained with help of the L-G functional, $S_{L-G}(\phi ),$
(which, in view of the Remark 3.2., can be identified with $\tilde{S}%
[\varphi ]$ defined by Eq.(3.7))\footnote{%
To be in accord with standard texts on statistical mechanics, e.g. see [17],
the constant scalar curvature $R(g)$ should be identified with the squared
"mass", i.e. with $m^{2}=a\left\vert T-T_{c}\right\vert $ where $T$ and $%
T_{c}$ are respectively the temperature and critical temperature. Also the
sign "-" in front of $\tilde{\lambda}$ term should be replaced by "+"} one
can develop things differently but, surely, equivalently.

To this purpose, instead of the functional $S[\varphi ]$ given in Eq.(3.2)
we use 
\begin{equation}
S_{V}[\varphi ]=\frac{1}{V^{\frac{2}{p}}}\int_{M}d^{d}x\sqrt{g}%
\{(\bigtriangledown _{g}\varphi )^{2}+R(g)\varphi ^{2}\}  \tag{6.23}
\end{equation}%
and replace $S[\varphi ]$ in the exponent of the path integral in Eq.(6.21)
by $S_{V}[\varphi ]$ from Eq.(6.23). Then, instead of Eq.(6.22), we obtain, 
\begin{equation}
\int D\left[ \phi \right] \exp \{-S_{L-G}(\phi )\}\ddot{=}%
\int\limits_{0}^{\infty }dVZ_{Y}(V)\text{,}  \tag{6.24}
\end{equation}%
where the sign \"{=} means \textquotedblright supposedly\textquotedblright .
This is so, because, in view of the Remark 3.3., at the level of saddle
point calculations the left hand side and the right hand side produce the
same G-L equation. Beyond the saddle point, calculations are not necessarily
the same. \ Although we plan to discuss this issue in subsequent
publications, some special cases are further discussed below in this section.

\textbf{Remark 6.3.} It should be clear, that at the level of the saddle
point calculation, replacement of the functional $S[\varphi ]$ in Eq.(6.21)
by $\mathcal{F}(\varphi )$ from Eq.(6.20) is completely adequate, so that
the sequence of steps in the analysis performed for the two dimensional case
in Section 5 are transferable to higher dimensions without change.

\textbf{Remark 6.4.} Although in 3 dimensions we have the result given by
Eq.(6.17), which forbids the use of identities like that in Eq.(5.6) still,
based on arguments just presented, the functional $\mathcal{F}(\varphi )$
defined by Eq.(6.20) should be used in the exponent of the corresponding
path integral replacing that given in Eq.(5.15) in 2 dimensions. Since by
doing so one will be confronted with the same type of minimization problems
as discussed earlier in Section 5\footnote{%
This conclusion was reached without any reliance on path integrals and on
physical applications in Ref.[51]. In this and related Ref.[20] the extremal
properties of determinants of $\square _{g}$ and $\Delta _{g}$ with respect
to variations of the background metric were studied.}, \ the functional
integral thus defined is an exact 3 dimensional analog of the path integral,
Eq.(5.30).

\ 

\subsubsection{\ \ \ A sketch of 3d CFT calculations}

Our obtained results allow us to proceed in complete accord with Section 5.
Thus, let $\hat{g}(u)=e^{\phi (u)}g$ be a 1-parameter family of metrics of
fixed volume and such that $\hat{g}(0)=g$. This implies that $\phi (0)=0$
and $\int\limits_{M}e^{\phi (u)}dV_{0}=V.$ In two dimensions, using results
of OPS [41], especially their Eq.(1.12), it is straightforward to obtain the
following result 
\begin{equation}
\frac{d}{du}(-\ln \det^{{}}\text{ }^{\prime }\Delta _{g(u)})\mid _{u=0}=\dot{%
\zeta}_{g}^{\prime }(0)=\frac{1}{12\pi }\int\limits_{M}\dot{\phi}K(g)dV_{g} 
\tag{6.25}
\end{equation}%
where $K(g)$ is the Gaussian curvature for the metric $g$. $\dot{\zeta}%
_{g}(0)$ and $\dot{\phi}$ represent $\frac{d}{du}\zeta _{g(u)}(0)\mid _{u=0}$%
and $\frac{d}{du}\phi (u)\mid _{u=0}$respectively. Volume conservation
implies$\ $%
\begin{equation}
\frac{d}{du}\int\limits_{M}e^{\phi (u)}dV_{g}\mid _{u=0}=\int\limits_{M}\dot{%
\phi}dV_{g}=0  \tag{6.26}
\end{equation}%
in accord with the earlier result, Eq.(5.17). If the Gaussian curvature $%
K(g) $ is constant, then Eq.(6.25) and Eq.(6.26) produce the same result.
This implies that $\frac{d}{du}\zeta _{g(u)}(0)\mid _{u=0}=0,$ that is $g$
is the \textquotedblright critical\textquotedblright\ (extremal) metric. In
view of Eq.(6.25) this also means that, for such a metric, the free energy
(e.g. see definition given by Eq.(5.5))\ attains its extremum. By combining
Eq.s (6.10), (6.14) and (6.15) we obtain at once 
\begin{eqnarray}
\delta _{f}\zeta _{\square _{g}}^{\prime }(0) &=&\delta
_{f}\int\limits_{0}^{\infty }dtt^{s-1}Tr(e^{-t\square _{g}})\mid _{s=0} 
\notag \\
&=&\int\limits_{0}^{\infty }dtt^{s}Tr(-2f\square _{g}e^{-t\square _{g}})\mid
_{s=0}.  \TCItag{6.27}
\end{eqnarray}%
From here we obtain essentially the same result as the main theorem by
Richardson [20], i.e. 
\begin{equation}
\dot{\zeta}_{\square _{\hat{g}}}^{\prime }(0)\mid
_{u=0}=0=\int\limits_{M}dV_{g}\dot{\phi}(x)\square _{g}\zeta (1,x,x). 
\tag{6.28}
\end{equation}%
That is, provided that we replace $\square _{g}$ by $\Delta _{g}$ and
require that \textit{locally} $\zeta (1,x,x)=const$, with the heat kernel $%
\zeta (s,x,x)$ given (as usual) by 
\begin{equation}
\zeta (s,x,x)=\sum\limits_{k=1}^{\infty }\frac{\psi _{k}^{2}(x)}{\lambda
_{k}^{s}},  \tag{6.29}
\end{equation}%
we obtain the main result by Richardson, Ref.[20], e.g. see his Theorem 1
and Corollary 1.1. Here $\psi _{k}(x)$ are eigenfunctions of the Laplacian
(or Yamabe Laplacian respectively) corresponding to eigenvalues $\lambda
_{k} $ with $0=\lambda _{0}<\lambda _{1}\leq \lambda _{2}\leq \cdot \cdot
\cdot \leq \lambda _{k}\leq \cdot \cdot \cdot .$ In the two dimensional
case, the condition for criticality: $\dot{\zeta}_{g}^{\prime }(0)=0$ is 
\textit{local} meaning that, provided the volume is constrained, the
constancy of the Gaussian curvature $K(g)$ at a given point of $M$ \ is
caused by the metric $g$ for which $\dot{\zeta}_{g}^{\prime }(0)$ is
extremal. In 3 and higher dimensions, constancy of the curvature at the
point of $M$ is being replaced by constancy of $\zeta (1,x,x)$ under the
same conditions of volume conservation. Such a condition is only necessary
but is now not sufficient, since the analog of Moser-Trudinger inequality,
Eq.(5.21), used to prove sufficiency in 2 dimensions does not exist.
Instead, one should study locally the second variation of $\zeta _{\hat{g}%
}^{\prime }(0)$ with respect to the underlying background metric in order to
decide if such a (local) extremum is a maximum or minimum. Fortunately, this
task was accomplished in Ref.s[20\textbf{, }51]. In particular, Richardson
[20] obtained the following theorem

\ 

\textbf{Theorem 6.5}.(Richardson) \textit{The Euclidean metric on a cubic
3-torus is a local maximum of determinant of the Laplacian with respect to
fixed-volume conformal variations of the metric.}

\ \ 

\textbf{Remark 6.6}. This Theorem is proven only for the cubic 3-torus. The
word \textquotedblright local\textquotedblright\ means that there could be
(or, there are, as we shall demonstrate) other 3-tori also providing local
maxima for determinants. In fact, according to the result by Chiu [52], all
determinants of flat 3-tori possess local maxima so that the determinant for
the face centered cubic (fcc) lattice has the largest determinant. Physical
implications of this fact can be found in our recent work, Ref.[47], and
also will be briefly considered in Section 8 below.

\bigskip

\textbf{Remark 6.7}. By using the same arguments as in Hawking's paper, Ref.
[53], the above theorem by Richardson can be given a physical
interpretation. It is based on use of the saddle point methods applied to
the Yamabe path integral, Eq.s (6.21), (6.24).

\ 

\bigskip The second variation of the Yamabe functional was calculated by
Muto [54] (see also Ref.[49]) with the result: 
\begin{equation}
\left( \frac{d^{2}}{dt^{2}}\mathcal{R}(g(t))\right) _{t=0}=\frac{d-2}{2}%
[\int\limits_{M}dV_{g}(\sigma (\bigtriangledown _{g}\varphi
)^{2}-R(g)\varphi ^{2})],  \tag{6.30}
\end{equation}%
where the constant $\sigma =d-1.$ As in the case of quadratic actions in the
flat space [17, 36] the second variation (with volume constrained to be
equal to one) looks very much the same as the original quadratic Yamabe
functional, except for the \textquotedblright wrong\textquotedblright\ sign
in front of the scalar curvature. This "wrong" sign has important physical
significance in both statistical mechanics and high energy physics. \ While
the significance to high energy physics will be discussed in the companion
publication, Ref.[32], here we would like to comment on the significance to
statistical mechanics. For this we need to recall that\ for the G-L free
energy functional $\mathcal{F}$\{$\varphi \}$ given by [17]%
\begin{equation}
\mathcal{F}\{\varphi \}=\mathcal{F}_{0}+\frac{1}{2}\int d\mathbf{x}\{c\left(
\nabla \varphi (\mathbf{x})\right) ^{2}+a\varphi ^{2}(\mathbf{x})+\frac{b}{2}%
\varphi ^{4}(\mathbf{x})\}  \tag{6.31}
\end{equation}%
the second variation is given by 
\begin{equation}
\mathcal{F}\{\varphi \}=\mathcal{F}\{\varphi _{0}\}+\frac{1}{2}\int dV\int
dV^{\prime }\left[ \frac{\delta ^{2}}{\delta \varphi (\mathbf{x})\delta
\varphi (\mathbf{x}^{\prime })}\mathcal{F}\{\varphi \}\mid _{\substack{ %
\varphi =\varphi _{0}  \\ h=0}}\right] \eta (\mathbf{x})\eta (\mathbf{x}%
^{\prime })+...,  \tag{6.32}
\end{equation}%
where%
\begin{equation}
\frac{\delta ^{2}}{\delta \varphi (\mathbf{x})\delta \varphi (\mathbf{x}%
^{\prime })}\mathcal{F}\{\varphi \}\mid _{\varphi =\varphi _{0}}=\left(
-c\bigtriangledown ^{2}+a+3b\varphi _{0}^{2}\right) \delta (\mathbf{x}-%
\mathbf{x}^{\prime }).  \tag{6.33}
\end{equation}%
with $\varphi _{0}$ being obtained from the extremum condition 
\begin{equation}
\frac{\delta \mathcal{F}\{\varphi \}}{\delta \varphi (\mathbf{x})}%
=-c\bigtriangledown ^{2}\varphi +a\varphi +b\varphi ^{3}-h=0  \tag{6.34}
\end{equation}%
In a special case of a coordinate independent field $\varphi _{0}$ Eq.(6.34)
is reduced to 
\begin{equation}
a\varphi _{0}+b\varphi _{0}^{3}=0.  \tag{6.35}
\end{equation}%
For $a<0$ $\ $we obtain : $\varphi _{0}=\pm (\left\vert a\right\vert /b)^{%
\frac{1}{2}}$ and $\varphi _{0}=0.$ The solution $\varphi _{0}=0$ is not a
minimum for the free energy and, hence, should be discarded. Of the two
other solutions, the system chooses one of them (which is interpreted in
literature as \textit{spontaneous symmetry breaking}). They both have the
same free energy\footnote{%
Once the choice is made, the order parameter can be considered in all
subsequent calculations as positive.}. If we let now $\varphi (\mathbf{x}%
)=\varphi _{0}+\delta \varphi \equiv \varphi _{0}+\eta (\mathbf{x})$ and
Taylor series expand $\mathcal{F}\{\varphi \}$ the result will coincide with
Eq.s(6.32) and (6.33). Notice that for $a<0$ a combination -$\left\vert
a\right\vert +3b\varphi _{0}^{2}=2\left\vert a\right\vert \equiv m^{2}$ (as
compared to $a$ in the high temperature phase). So, if in the high
temperature phase $m^{2}=R(g)$ with $R(g)>0,$ in the low temperature phase
(i.e. below the criticality) $R(g)<0!$ Moreover, by choosing $d=4$ in
Eq.(6.30) we obtain a combination: $m^{2}=\frac{4-2}{2}R(g)=2R(g)$ in accord
with the standard G-L results. The constant c in Eq.(6.33) is left
unspecified in G-L theory while now it is equal to 6 (for $d=4).$The spaces
of constant negative curvature are hyperbolic.

\textbf{Remark 6.8. }In our earlier work, Ref[55] we emphasized the major
importance of hyperbolicity in statistical mechanical calculations related
to the AdS/CFT correspondence. In this work we arrived at the same
conclusions quite independently.

From the work by Muto [54] we know that: a) if $R(g)$ is positive, the
second variation can be made positive for appropriately chosen $\varphi 
\footnote{%
This can be easily understood if we expand $\varphi $ into Fourier series
made of eigenfunctions of the Laplacian and take into account that for 
\textbf{any} closed Riemannian manifold the spectrum of the Laplacian is
nonnegative and nondecreasing [29].},b)$ if $R(g)$ is negative, the second
variation is positive for the same reasons. The positivity of the second
variation implies that the extremal \textit{constant curvature} metric $g$
provides a \textit{locally stable} \textit{minimum }for $\mathcal{R}(g(t))$
\ defined by Eq.(4.9). That is, \ using results of Section 4, the Einstein
metric obtained as solution \ to Eq.(4.2) is stable among nearby metrics.
This conclusion will be reinforced in the next (dynamical) section.

\ 

\textbf{Remark 6.9}. It is interesting to notice that the calculation of
higher order fluctuation corrections to the Yamabe path integral, Eq.(6.21),
involves calculations on the moduli space of Einsteinian metrics, Ref.[56].
This observation provides a strong link between higher dimensional LGW
theory and two dimensional string inspired CFT's discussed in the previous
section. Naturally, Eq.(6.24) can be used to investigate to what extent the
final results of conventional field-theoretic calculations, Ref.[17,36],
might differ from more sophisticated string-theoretic calculations in the
style of Ref.s[14,15,46].Additional details can also be found in our work,
Ref.[47].

\section{Critical dynamics and the Yamabe \ and Ricci flows}

\subsection{ Physical motivation}

\textit{\ }The results obtained in previous sections are somewhat
unrealistic from a purely practical (physical) point of view. The situation
in our case is analogous to that known in thermodynamics. Recall, that this
discipline emerged from practical needs to improve the efficiency of heat
engines. Clearly, rigorously speaking, it is not applicable to such devices
since, by definition, it is valid only for time-independent phenomena.
Hence, it cannot be used at all because it usually takes a very long (if not
infinite) time for the system to equilibrate, especially near the critical
temperature $T_{c}$. Thus, the description of phase transitions by the G-L
theory is valid only in a rather narrow range of temperatures around $T_{c}$%
. To bring a physical system into this range of temperatures (under constant
pressure) requires varying the temperature of the surrounding environment in
time. In addition,\ by definition, the \textquotedblright
true\textquotedblright phase transition should take place only in the
thermodynamic limit (of infinite volume with particle density kept
constant). Since in the real world the systems under study are always of
finite size, this requirement is implemented by imposing physically
appropriate boundary conditions, e.g. periodic. Use of some appropriate
boundary conditions causes systems undergoing phase transitions to actually
"live" on some manifolds/orbifolds. Under such circumstances the topology
and physics become intertwined. The signature of such boundary effects can
be seen already in the calculation of determinants discussed in Sections 5
and 6. More on this subject is discussed in our earlier work, Ref.[55],
where, in accord with Remark 6.8., we emphasized the role of hyperbolic
spaces in the theory of phase transitions\footnote{%
Remark 6.8. is not in contradiction with the remark 6.6. This is explained
in great detail in our recent work, Ref.[47] and, is also discussed in
Section 8.}.

The equilibration process known in the physics literature as \textit{%
critical dynamics} can take a very long time (infinite in the thermodynamic
limit for $T\Longrightarrow T_{c})$. Since idealization of reality is
typical in physics, \ in this work we adopt the pragmatic (physical) point
of view by considering systems of finite size with appropriately chosen
boundary conditions. Then, the results of the previous sections are limiting
cases of more general time-dependent G-L theory considered for such systems.

Development of time-dependent G-L theory was initiated by Landau and
Khalatnikov in 1954 [57] and is also phenomenological. As such, it is based
on the assumption that an order parameter $\varphi $ satisfies the
relaxation equation of the type given by 
\begin{equation}
\frac{\partial \varphi }{\partial t}=-\gamma \frac{\delta \mathcal{F}%
\{\varphi \}}{\delta \varphi }\equiv -\gamma grad\mathcal{F}\{\varphi \} 
\tag{7.1}
\end{equation}%
with functional $\mathcal{F}\{\varphi \}$ defined by Eq.(6.31) and the
\textquotedblright friction\textquotedblright\ coefficient $\gamma $ is some
(assumed to be) known non-negative constant. By rescaling time it can be
eliminated. Such rescaling is assumed in all calculations below. Since such
an equation was postulated, its validity was checked by real experiments
with excellent outcome [58\textbf{]}. Being armed with such results, we
would like to develop the mathematical formalism of the previous sections to
account for the effects of critical dynamics.

\subsection{Mathematical motivation \textit{\ }}

\bigskip

Eq.(7.1) is an example of the gradient flow. In this subsection we would
like to place the Landau-Khalatnikov theory of critical dynamics on a more
rigorous mathematical basis also involving the notion of the gradient flow.

We begin with reviewing the results of OPS, Ref. [41], (obtained for two
dimensional manifolds without boundaries) in the style of Hamilton's work,
Ref.[16]. To this purpose we notice that Eq.(5.19) is an extremum of the
functional $\mathcal{F}(\psi )$, Eq.(5.14). Clearly, if we write $\varphi
=\psi +\varepsilon h$ in this functional so that 
\begin{equation}
\frac{d}{d\varepsilon }\mathcal{F}(\psi +\varepsilon h)\mid _{\varepsilon
=0}=-\int\limits_{M}dA_{g}\left( \Delta _{g}\psi \right) h+\frac{1}{2}%
\int\limits_{M}dA_{g}R(g)h-2\pi \chi (M)\frac{\int\limits_{M}dA_{g}h\exp
(2\psi )}{\int\limits_{M}dA_{g}\exp (2\psi )}  \tag{7.2}
\end{equation}%
then, following OSP, the scalar product $\left\langle ,\right\rangle $ in
the tangent space to each point at which $\psi $ is defined can be
introduced according to convention: 
\begin{equation}
\left\langle grad\mathcal{F},h\right\rangle \equiv \left( \delta \mathcal{F}%
\right) h=\frac{d}{d\varepsilon }\mathcal{F}(\psi +\varepsilon h)\mid
_{\varepsilon =0}.  \tag{7.3}
\end{equation}%
In accord with Eq.(3.6) by OPS, Ref.[41], this convention implies that 
\begin{equation}
grad\mathcal{F\{\psi \}=}-\Delta _{g}\psi +\frac{1}{2}R(g)-\frac{2\pi \chi
(M)\exp (2\psi )}{\int\limits_{M}\exp (2\psi )dA_{g}}.  \tag{7.4}
\end{equation}%
This result is to be compared with Eq.(5.19). Naturally, Eq.(5.19)
corresponds to the condition of equilibrium, i.e. $grad\mathcal{F\{\psi \}}%
=0.$ Away from equilibrium the dynamical equation \textit{postulated} by OPS
(see below) \textit{coincides} with Eq.(7.1) postulated much earlier by
Landau and Khalatnikov. To see this, \ we need to take into account Eq.s
(5.2), (5.19) and (5.20).With their help we would like to rewrite Eq.(7.4)
in the following \textit{equivalent} form, 
\begin{equation}
-grad\mathcal{F\{\psi \}=}\frac{1}{2}\mathcal{(}<\hat{R}(\hat{g})>-\hat{R}(%
\hat{g}))e^{2\psi },  \tag{7.5}
\end{equation}%
where $<\hat{R}(\hat{g})>=\left( \int\limits_{M}d\hat{A}\hat{R}(\hat{g}%
)\right) /\int\limits_{M}d\hat{A}$ with \ $d\hat{A}$ given by Eq.(5.3).\
Accordingly, \ in view of Eq.(7.1), \textit{up to time rescaling}, the
equation describing\ critical dynamics can be written as follows%
\begin{equation}
\frac{\partial \psi }{\partial t}=<R(g)>-R(g)  \tag{7.6}
\end{equation}%
coinciding with Eq.(3.9) of OPS\footnote{%
In making such a comparison (as in Section 5) we took into account that $%
R=2K $.}

\bigskip Suppose that the equilibrium (the fixed point) condition, 
\begin{equation}
<R(g)>-R(g)=0,  \tag{7.7}
\end{equation}%
is valid in dimensions higher than two as well. Then it coincides formally
with Eq.(2.12) defining scalar curvature or with Eq.(4.12) where it was
obtained from the Hilbert-Einstein functional for pure gravity in the
presence of the cosmological term. \ In view of this observation, we would
like to demonstrate that equations like \ Eq.(7.6) whose fixed points are
determined by the\ equations of the type given by Eq.(7.7)\ indeed can be
obtained in higher dimensions. In this work only 3 dimensional results will
be discussed in some detail.

\subsection{\textit{\ }Critical dynamics in dimensions higher than two:
Yamabe vs Ricci flows\textit{\ }}

\bigskip\ Following Hamilton [16], we begin with

\ 

\textbf{Definition 7.1}\textit{.} The \ normalized \textit{Ricci flow} is
described by the dynamical equation given by 
\begin{equation}
\frac{\partial }{\partial t}g_{ij}=\frac{2}{d}g_{ij}<R(g)>-2R_{ij}(g). 
\tag{7.8}
\end{equation}

\bigskip\ 

In writing this equation we use the notations of Section 4 where we defined $%
<R(g)>=\left( \int_{M}R\sqrt{g}d^{d}x\right) /\int_{M}\sqrt{g}d^{d}x\equiv
\left( \int_{M}Rd\mu \right) /\int_{M}d\mu .$ The above flow equation
formally exists for $d\geq 2$ and should be considered along with some
prescribed initial condition: $\ g_{ij}(t=0)=\hat{g}_{ij}.$ In two
dimensions it is always possible to write $R_{ij}(g)=\frac{1}{2}Rg_{ij\text{ 
}}$ so that all two dimensional spaces are Einsteinian [16]. Because of this
fact, Eq.(7.8) is converted into 
\begin{equation}
\frac{\partial }{\partial t}g_{ij}=(<R(g)>-R(g))g_{ij}\text{ , }d=2. 
\tag{7.9}
\end{equation}

\textbf{Definition 7.2.} Eq.(7.9) defines the \textit{Yamabe flow}. As such,
it can be considered for any $d\geq 2.$

\textbf{Remark 7.3.} If, like in Section 2, we choose $\hat{g}_{ij}=e^{\psi
}g_{ij}(0)$ and substitute this result into Eq.(7.9), we reobtain Eq.(7.6).
Hence, Eq.(7.6) does describe the Yamabe flow.

\textbf{Corollary 7.4.} Using results of previous subsection (at least in
two dimensions) the Yamabe flow is the \textit{gradient} flow.

\bigskip

\textbf{Corollary 7.5.}Since the Yamabe\textbf{\ }flow in Eq.(7.9)\textbf{\ }%
is defined for any $d\geq 2,$ the fixed points for such a flow should
coincide with those given by Eq.(7.7).

\ 

\textbf{Remark 7.6. }Only in 2\textbf{\ }dimensions are the Yamabe and the
normalized Ricci flows (essentially) equivalent. This is explained in great
detail by Hamilton [16]. \ Since results of this paper crucially depend on
the observation that \textit{only} in 2 dimensions it is \textit{always}
possible to write $R_{ij}(g)=\frac{1}{2}Rg_{ij},$ it should be clear that
attempts to extend this equivalence to higher dimensions are destined for
failure in those cases where the flow takes place on \ spaces which are not
of Einstein type. Hence, in this case the task lies in finding conditions
under which the Ricci flow, for which the initial metric is chosen to be of
not necessarily Einstein-type, leads to an Einstein-type metric(s) as the
fixed point(s) solution(s) for such a flow. This is discussed in the next
subsection.

\bigskip

Even though Yamabe and Ricci flows are different in higher dimensions, one
may still ask a related question : can one use some Ricci flows in order to
obtain results for Yamabe -type flows? We would like to demonstrate that
this is indeed possible. To this purpose, using Eq.(7.9) we obtain, 
\begin{equation}
\frac{1}{2}g^{ij}\frac{\partial }{\partial t}g_{ij}=<R(g)>-R(g).  \tag{7.10}
\end{equation}%
In addition, since in any dimension $\frac{\partial }{\partial t}d\mu \equiv 
\frac{\partial }{\partial t}\sqrt{\det g_{ij}}d^{d}x=\frac{1}{2}g^{ij}\frac{%
\partial }{\partial t}g_{ij}d\mu $ $,$ we obtain as well 
\begin{equation}
\frac{\partial }{\partial t}\tint\nolimits_{M}d\mu =\tint\nolimits_{M}\frac{%
\partial }{\partial t}d\mu =\tint\nolimits_{M}(<R>-R(g))d\mu =0.  \tag{7.11}
\end{equation}%
This equation is compatible with Eq.(7.8) and valid for $d\geq 2$. Indeed,
if we take into account that $\frac{\partial }{\partial t}d\mu =\frac{1}{2}%
g^{ij}\frac{\partial }{\partial t}g_{ij}d\mu ,$ use Eq.(7.8) for $\frac{%
\partial }{\partial t}g_{ij}$, multiply both sides of this equation by $%
g^{ij}d\mu $ and integrate the obtained result over $M$, we reobtain
Eq.(7.11). Hence, after some manipulations Eq.(7.8) describing Ricci flow
(and \ valid for $d\geq 2$ ) can be brought into a form identical with \
Eq.(7.10) describing Yamabe flow (obtained originally for $d=2)$.This
observation does not imply that both flows are equivalent as Remark 7.6.
indicates. Nevertheless, the Yamabe flow can be looked upon as some special
case of the Ricci flow.

\textbf{Corolary 7.7}. The arguments just presented demonstrate that
solutions for the Yamabe flow can be obtained from solutions for the
normalized Ricci flow. Moreover, in both cases the fixed points for such
flows are given by Eq.(7.7) thus providing justification for the
compatibility of 2 and 3 dimensional CFT discussed in previous sections and,
hence, for the unique pathway for extending the results of 2d CFT to 3 and
higher dimensions.

\textbf{Corollary 7.8}. Since the fixed point, Eq.(7.7), is the traced form
of the fixed point equation for the Ricci flow, Eq.(7.8), that is of the
equation 
\begin{equation}
\frac{1}{d}g_{ij}<R(g)>-R_{ij}(g)=0,  \tag{7.12}
\end{equation}%
which is equivalent to Eq.(4.2), this means: a) that the fixed point
solutions of the normalized Ricci flow always produce the Einstein spaces,
provided that these fixed points are stable and, b) that the arguments
presented in Remark 4.3. are indeed valid.

\subsection{Stability of the fixed points solutions}

The results obtained previously in this section cannot be used until the
stability analysis of these solutions is performed. Since according to the
Corollary 7.5. solutions for the Yamabe flow can be obtained from those for
the Ricci flow, it is sufficient, in principle, to consider only the
stability of the fixed point solutions for the Ricci flows. This path is not
the most physically illuminating however as we would like to explain now.

\subsubsection{ Dynamics of the Yamabe flows}

\bigskip The Yamabe flow described by Eq.(7.9) should be supplemented by the
initial condition, e.g. $g_{ij}(t=0)=g_{ij}(0)\equiv \hat{g}.$ In order to
study the evolution of this metric it is convenient to write it in the
following form [59]: $g_{ij}(t)=\left[ \varphi (t)\right] ^{\frac{4}{d-2}%
}g_{ij}(0).$Using such a substitution in Eq.(7.9) (along with Eq.(2.12))
allows us to rewrite it in the following equivalent form%
\begin{equation}
\frac{\partial \varphi }{\partial t}=-\varphi ^{2-p}(\alpha \Delta _{\hat{g}%
}\varphi +R(\hat{g})\varphi )+<R(g)>\varphi  \tag{7.13}
\end{equation}%
to be compared with the phenomenological result, Eq(7.1), by Landau and
Khalatnikov. Although these equations look somewhat different, their fixed
points (if these are stable) are the same \footnote{%
In two and three dimensions these fixed points were discussed in previous
sections.}. In view of the results of previous sections, we believe that the
equations of critical dynamics known in the physical literature, e.g. see
Ref.[58], all \ having their origin in the work by Landau and Khalatnikov
[57], should be replaced by Eq.(7.13) also describing critical dynamics of
physical systems of the Ginzburg-Landau type. Such a replacement is in
accord with the requirement of conformal invariance at criticality as
described in detail in our earlier work on AdS/CFT correspondence [55\textbf{%
]}.

\textbf{Remark 7.9}. Evidently, physical realization of the Yamabe and Ricci
flows\footnote{%
And, ultimately, the physically assisted proof of the Poincar$e^{\prime }$
and geometrization conjectures depend crucially on such an assumption.}
depends crucially on the possibility of such a replacement and, in view of
results of this work, it can be considered as proven.

Using Eq.(7.10) we would like to introduce the following quantity%
\begin{equation}
\eta =\frac{4}{d-2}(<R(g)>-R(g)).  \tag{7.14}
\end{equation}%
In view of Eq.(7.11), integrating Eq.(7.14) we obtain,%
\begin{equation}
\tint\nolimits_{M}\eta d\mu =0.  \tag{7.15}
\end{equation}%
By combining Eq.s(7.11) and (7.13), after some calculations which can be
found in Ref.[59], we obtain,%
\begin{equation}
\frac{\partial }{\partial t}<R(g(t))>=\frac{d-2}{2}\tint\nolimits_{M}R(g)%
\eta d\mu .  \tag{7.16}
\end{equation}%
Next, by combining Eq.s(7.14) and (7.16) we further obtain,%
\begin{equation}
\frac{\partial }{\partial t}<R(g(t)))>=2\tint%
\nolimits_{M}R(g)(<R(g)>-R(g))=-2\tint\nolimits_{M}(<R(g)>-R(g))^{2}, 
\tag{7.17}
\end{equation}%
where the second equality comes from the constraint, Eq.(7.15). The last
expression possesses an entropic meaning. Indeed, following Boltzmann [60],
we introduce the $H$-function (the entropy) for some gas made of hard
spheres elastically scattering from each other . Explicitly, it is given by 
\begin{equation}
H=-\tint f\ln fd\mathbf{v,}  \tag{7.18}
\end{equation}%
where the distribution function $f(\mathbf{v,}t)$ obeys the Bolzmann's
equation%
\begin{equation}
\frac{\partial }{\partial t}f=\tint (f(\mathbf{v}^{\prime }\mathbf{,}t)f(%
\mathbf{v}_{1}^{\prime }\mathbf{,}t)-f(\mathbf{v,}t)f(\mathbf{v}_{1}^{{}}%
\mathbf{,}t))\left\vert \mathbf{v}_{1}-\mathbf{v}\right\vert I(\theta
,\left\vert \mathbf{v}_{1}-\mathbf{v}\right\vert )d\mathbf{v}_{1}d\Omega 
\tag{7.19}
\end{equation}%
with $I(\theta ,\left\vert \mathbf{v}_{1}-\mathbf{v}\right\vert )$ being
some known function, $d\Omega =\sin \theta d\theta d\phi ,$ and $\mathbf{v}%
^{\prime }=\mathbf{v}+\mathbf{n}(\mathbf{n}\cdot \mathbf{g}),\mathbf{v}%
_{1}^{\prime }=\mathbf{v}_{1}-\mathbf{n}(\mathbf{n}\cdot \mathbf{g})$ being
respective velocities of the colliding particles after (primed) and before
(nonprimed) scattering on each other$,$ here $\mathbf{g=v}_{1}-\mathbf{v,}$
and the vector $\mathbf{n}$ is a unit normal\footnote{%
See \textbf{[}60] for details.}. Based on these results, it can be shown
that $d\mathbf{v}d\mathbf{v}_{1}=d\mathbf{v}^{\prime }d\mathbf{v}%
_{1}^{\prime }.$ Using these results, let us consider calculation of $\frac{%
\partial }{\partial t}H.$By combining Eq.s(7.18) and (7.19) one obtains
after some calculation $\frac{\partial }{\partial t}H\geq 0$ with an
equality only for the equilibrium case for which $f^{\ast }(\mathbf{v}%
^{\prime }\mathbf{,}t)f^{\ast }(\mathbf{v}_{1}^{\prime }\mathbf{,}t)-f^{\ast
}(\mathbf{v,}t)f^{\ast }(\mathbf{v}_{1}^{{}}\mathbf{,}t)=0.$ This condition
is leading to the Maxwell's distribution of particle velocities, which
produce a well tested equation of state for the ideal gas. In accord with
thermodynamics, such defined entropy riches its maximum at equilibrium.

\textbf{Remark 7.10. }The\textbf{\ }obtained equilibrium result for
Maxwell's distribution depends crucially on the assumption of occurrence of
only binary collisions in the gas. Boltzmann's equation, Eq.(7.19), reflects
just this fact. The obtained results become much less tractable if one would
like to account for ternary and higher order collisions. The same is true if
one uses the full renormalization group flow equations [4] instead of just
the leading order terms in these equations. The proof of the Bolzmann's
H-theorem as well as the proof of the Poincar$e^{\prime }$ and
geometrization conjectures are based on the validity of such approximations.
The validity is assured by excellent agreement with experiment in
Boltzmann's case and with equally good agreement with experiment in the case
of Landau-Ginzburg critical dynamics described by the Yamabe-type flow.

Based on these physical arguments, we may choose $-<R(g(t)))>$ as our
entropy (provided that the volume of the system is normalized, say, to
unity). Then, we can identify Eq.(7.9) (or (7.13)) with a Boltzmann-type
equation. Our previously obtained result, Eq.(4.12), then provides the
physically meaningful equilibrium solution. Moreover, using these results we
arrive at the conclusion that the fixed point solutions for the Yamabe flow
are described by the Einstein equation, Eq.(4.2), leading to the Einstein
manifolds of constant scalar curvature in accord with the Remark 4.3.

\subsubsection{\protect\bigskip Dynamics of the Ricci flows}

In this subsection we would like to demonstrate in some detail that, indeed,
some Ricci flows produce results that are in agreement with those obtained
for the Yamabe flows. For this purpose, following the logic of the previous
subsection we have to find an analog of the entropy for the Ricci flows.
This task was accomplished by Perelman in the first of his 3 papers, Ref.[1%
\textbf{]}. He noticed, albeit indirectly, that, unlike the Yamabe flow
(which is the gradient-type flow), the Ricci flow is not gradient. The proof
can be found in Ref.[61]. Accordingly, one cannot find an entropy for such a
flow and, hence, to study its stability. Perelman found ingenious way around
these difficulties. In view of the existing literature [2,3, 10, 61-63], our
exposition will be brief emphasizing the physical aspects of the results he
had obtained\footnote{%
Naturally, only small part of Perelman's results is discussed in this paper.}%
.

\paragraph{The Boltzmann-Nash entropy and \ the dilaton gravity.}

\ In the case of the Yamabe flow our choice of $-<R(g(t)))>$ as an entropy
seemed somewhat artificial in the sense that it was not motivated by any
systematic procedure involving entropy as such. We would like to correct
this deficiency in this subsection.We begin with the following

\textbf{Definition 7.11}.The heat operator $\square :=\partial _{t}-\Delta
_{g}$ has an operator $\square ^{\ast }=-\partial _{t}-\Delta _{g}+R(g)$ as
its adjoint with respect to the Ricci flow. E.g. see Ref.[61], page 22.

\bigskip \textbf{Remark 7.12}. The proof is based on the observation that
upon some rescaling [5] it is possible to supress the normalization in the
Ricci flow, Eq.(7.8), that is to eliminate the factor $\frac{2}{d}%
g_{ij}<R(g)>$. As result of this elimination, Eq.(7.10) acquires the form%
\begin{equation}
\frac{1}{2}g^{ij}\frac{\partial }{\partial t}g_{ij}=-R(g)  \tag{7.20}
\end{equation}%
used in the proof of the conjugacy of \ operators $\square $ and $\square
^{\ast }.$

Let low $g_{ij}$ be a solution of the non normalized Ricci flow equation and 
$u=exp(-f)$ be the solution of the equation $\square ^{\ast }u=0$ for some
function $f$ such that $.\tint\nolimits_{M}e^{-f}$ $dV=1$. By nanalogy with
Boltzmann's entropy, Eq.(7.18), following Ref.[61] we define the Nash
entropy $N(u)$ as follows:%
\begin{equation}
N(u)=\tint\nolimits_{M}u\ln u\text{ }dV.  \tag{7.21}
\end{equation}%
The time derivative $\partial _{t}N(u)$ \ can be calculated now with help of
\ the adjoint equation $\square ^{\ast }u=0.$ Indeed, we obtain:%
\begin{eqnarray}
\partial _{t}N(u) &=&\tint\nolimits_{M}(\partial _{t}u\ln u\text{ }%
dV+\partial _{t}u\text{ }dV\text{ }+u\ln u\text{ }\partial _{t}dV)  \notag \\
&=&\tint\nolimits_{M}((\partial _{t}-R(g))u\ln u+\partial _{t}u\text{ )}dV 
\notag \\
&=&\tint\nolimits_{M}(-\Delta _{g}u\ln u+(R(g)-\Delta _{g})u)dV. 
\TCItag{7.22a}
\end{eqnarray}%
On a closed manifold, the integral of $\Delta _{g}u$ vanishes. Because of
this, the last result acquires the form%
\begin{equation}
\partial _{t}N(u)=\tint\nolimits_{M}(\frac{\left\vert \bigtriangledown
_{g}u\right\vert ^{2}}{u}+R(g)u)dV=\tint\nolimits_{M}(\left\vert
\bigtriangledown _{g}f\right\vert +R(g))e^{-f}dV\equiv \mathcal{F}(g_{ij},f)
\tag{7.22b}
\end{equation}%
The functional $\mathcal{F}(g_{ij},f)$ is Perelman's entropy [1] and is also
an action for the dilaton gravity [ 8,64]. When $f$ is a constant, taking
into account the sign differences between the Bolzmann's entropy, Eq.(7.18),
and the Nash entropy, Eq.(7.21), we obtain back our earlier result for the
entropy $-<R(g(t)))>$ as required.

\textbf{Remark 7.13.} The entropic nature of the Einstein-Hilbert \ and
dilaton gravity actions just described might be especially useful for
applications to black hole thermodynamics. Indeed, the attempts to use the
Ricci flow for the description of black hole dynamics were recently made in
Ref.[65]. We plan to return to this issue in subsequent publications.

\textbf{Remark 7.14. }If one starts\textbf{\ }with the functional\textbf{\ }$%
\mathcal{F}(g_{ij},f)$ without prior remarks on its entropic origin, one
might employ the dilaton field $f$ which is time-independent. This leads to
some technical simplifications to be discussed below.

Having defined the entropy for the Ricci flow, the task now lies in
obtaining the analog of Eq.(7.17). We would like to do so in a way
consistent with our results for the static case.

\paragraph{Steady Solitons.}

In particular, following \ the logic of Section 7.2., we should begin with
the replacement of the two dimensional functional $\mathcal{F}(\psi )$ by
its three dimensional analog, e.g. by \~{S}[$\varphi ]$ defined by Eq.(3.7),
or, equivalently, by S[$\varphi ]$ defined by Eq.(3.2). To make a connection
with Perelman's work, we need temporarily to make a short but important
detour. To this purpose we set $\varphi =e^{-f/2}$ in Eq.(7.22). This causes
the functional $\mathcal{F}(g_{ij},f)$ to acquire the following new look 
\begin{equation}
\mathcal{F}(g_{ij},\varphi )=\tint\nolimits_{M}(4\left\vert \nabla \varphi
\right\vert ^{2}+R(g)\varphi ^{2})dV  \tag{7.23}
\end{equation}%
to be compared with the functional $E[\varphi ]$ defined in Eq.(3.2). By
analogy with Eq.(3.4), now we can define the constant $\lambda _{g}$ in
terms of the Raleigh quotient, i.e.%
\begin{equation}
\lambda _{g}=\inf_{\varphi }\frac{\mathcal{F}(g_{ij},\varphi )}{%
\tint\nolimits_{M}\varphi ^{2}dV}.  \tag{7.24}
\end{equation}%
In accord with Eq.(3.3), the constant $\lambda _{g}$ serves as an eigenvalue
in the equation\footnote{%
See the second footnote for the sign convention for the Laplacian.}%
\begin{equation}
4\Delta _{g}\bar{\varphi}+R(g)\bar{\varphi}-\lambda _{g}\bar{\varphi}=0, 
\tag{7.25}
\end{equation}%
where $\bar{\varphi}$ is the minimizer for the Raleigh quotient, Eq.(7.24).
Equivalently,%
\begin{equation}
\lambda _{g}=\inf \{\tint\nolimits_{M}(4\left\vert \nabla \varphi
\right\vert ^{2}+R(g)\varphi ^{2})dV,\text{ }\tint\nolimits_{M}\varphi
^{2}dV=1\}.  \tag{7.26}
\end{equation}%
By analogy with Eq.s (4.7)-(4.10),\ \ we introduce the family of metrics $%
g_{ij}(s)=g_{ij}+s$ $h_{ij}$ in order to consider the Raleigh quotient under
such a variation. Then, instead of Eq.(4.10), we obtain\footnote{%
For details of such calculations, please, consult [10] and [66].} 
\begin{equation}
\frac{d}{ds}\lambda (g_{ij}(s))=\tint\nolimits_{M}(-h_{ij})(R_{ij}+\nabla
_{i}\nabla _{j}f)e^{-f}dV.  \tag{7.27}
\end{equation}%
Since this result is analogous to that given in Eq.(7.16), the course of
action is going to be the same. In particular, following Perelman, and also
[10], we notice that the variation of the functional $\mathcal{F}(g_{ij},f)$
leads to the following coupled equations for the modified Ricci flow%
\footnote{%
Please, see again the sign convention for the Laplacian in the second
footnote.}%
\begin{equation}
\frac{\partial }{\partial t}g_{ij}=-2(R_{ij}+\nabla _{i}\nabla _{j}f), 
\tag{7.28a}
\end{equation}%
\begin{equation}
\frac{\partial }{\partial t}f=-R+\Delta f.  \tag{4.28b}
\end{equation}

\textbf{Definition 7.15}. The flow defined by these coupled equations is
called the \textit{generalized Ricci flow}.

Finally, after making the identification $\frac{\partial }{\partial t}%
g_{ij}=h_{ij}$ and using Eq.(7.28a) in Eq.(7.27) we obtain the monotonicity
result of Perelman%
\begin{equation}
\frac{d}{ds}\lambda (g_{ij}(s))=2\tint\nolimits_{M}\left\vert R_{ij}+\nabla
_{i}\nabla _{j}f\right\vert ^{2}e^{-f}dV  \tag{7.29}
\end{equation}%
to be compared with Eq.(7.17). In view of this comparison, we conclude that $%
\frac{d}{ds}\lambda (g_{ij}(s))=0$ \ only in the case when 
\begin{equation}
R_{ij}+\nabla _{i}\nabla _{j}f=0.  \tag{7.30}
\end{equation}

\textbf{Definition7.16.} In the existing terminology Eq.(7.30) is known as
the equation for the \textit{gradient} \textit{steady soliton} [ 1-3, 66].

Following Perelman [part I] we can extend this result to the case of the so
called breathers.

\textbf{Definition 7.17}. A metric $g_{ij}(t)$ evolving by the Ricci flow is
called a \textit{breather}, if for some $t_{1}<t_{2}$ and $\alpha >0$ the
metric $\alpha g_{ij}(t_{1})$ and $g_{ij}(t_{2})$ differ only by a
diffeomorphism. The cases when $\alpha =1,\alpha <1$ and $\alpha >1$
correspond to \textit{steady}, \textit{shrinking} and \textit{expanding}
breathers respectively. \textit{Solitons are trivial breathers} for which $\ 
$the above relationship between metrics holds for \textit{any} pair of $%
t_{1} $ and $t_{2}$ .

\textbf{Corollary 7.18. }If one considers the Ricci flow as a dynamical
system on the space of Riemannian metrics modulo diffeomorphisms, then the
breathers and the solitons are respectively the periodic orbits and fixed
points for such a system.

\textbf{Corollary 7.19}. In view of the result, Eq.(7.29), one can write $%
\lambda (g_{ij}(s_{1}))\leq \lambda (g_{ij}(s_{2})).$Because of Eq.(7.30) we
obtain as well: $\lambda (g_{ij}(s_{1}))=\lambda (g_{ij}(s_{2}))$ $\forall
s\in \lbrack s_{1},s_{2}].$ Hence, Eq.(7.30) providing \ a minimum for $%
\lambda (g_{ij}(s))$ is a steady breather and a steady soliton at the same
time. That is, there are no breathers for the Ricci flows in accord with
Perelman [1].

\ \ \ \ \ Going back to Eq.(7.30), multiplying it by $g^{ij}$ and making a
summation over repeated indices we obtain,%
\begin{equation}
R=\Delta _{g}f,  \tag{7.31}
\end{equation}%
where, \ as before, we used the fact that $-\Delta _{g}f=\nabla ^{i}\nabla
_{i}f.$Taking into account arguments which lead from Eq.(7.22a) to (7.22b)
we conclude that for compact manifolds $f=const$ and, hence, $R=0$.

\paragraph{Expanding and shrinking solitons.}

The obtained result is a special case of more general result by Ivey [67]
who extended the work by Hamilton [16] for surface to Ricci flows on compact
3- manifolds\footnote{%
Later, in Ref.[68], Hamilton came up with yet another proof of Ivey's
results.}. It is instructive to conclude this subsection by discussing the
physical significance of their results, especially in connection with
results obtained in the previous section.

To this purpose, let us recall that the fundamental solution of the heat
equation in $d-$dimensional Euclidean space can be written as [63]%
\begin{equation}
u(x,y,\tau )=\left( 4\pi \tau \right) ^{-\frac{d}{2}}\exp (-\left\vert
x-y\right\vert ^{2}/4\tau ),  \tag{7.32}
\end{equation}%
where $\tau =t-T$ or $T-t=-\tau \equiv \sigma $ depending upon wether we are
dealing with the forward or the backward heat equation\footnote{%
Here $T$ is some pre assigned time.}. If we let $y=0$, then such a solution
can be interpreted as probability\footnote{%
The fundamental solution, E.q.(7.32), also has the meaning of probability in
polymer physics [69]. This fact is briefly discussed in Section 8.} since by
design $\ $we have $\tint\nolimits_{M}udV=1.$ Using this fact, one can look
for a solution of the heat equation on some Riemannian manifold $M$ by
employing the ansatz $u=\left( 4\pi \tau \right) ^{-\frac{d}{2}}e^{-f}.$
This amounts of redefining the earlier introduced function $f:f=\tilde{f}-%
\frac{d}{2}\ln (4\pi \tau )$ so that $e^{-\tilde{f}}=\left( 4\pi \tau
\right) ^{-\frac{d}{2}}e^{-f}.$ Evidently, $\nabla f=\nabla \tilde{f}$ and,
therefore, \ $\Delta _{g}f=\Delta _{g}\tilde{f}$ as well. To simplify
matters, we would like to consider how things are done in the flat case.
Using Eq.(7.32) we begin with the calculation of the Nash entropy. A simple
calculation produces: $N_{flat}=-\frac{d}{2}-\frac{d}{2}\ln (4\pi \tau ).$
Using this result, we now define the properly normalized Nash entropy, i.e. $%
N(u)-N_{flat}$. Explicitly, we obtain,%
\begin{equation}
\tilde{N}(u)=N(u)-N_{flat}=\tint\nolimits_{M}(-f+\frac{d}{2})udV.  \tag{7.33}
\end{equation}%
Perelman \ does not explain how he had obtained his result in his work. He
calls such a normalized Nash entropy a "partition function", e.g. see
Section 5 of Part I of Ref.[1]. This is a bit misleading however because in
the same section he earlier defines the partition function correctly, i.e.
in accord with the accepted rules of statistical mechanics. According to
these rules one defines the \textit{free energy} $\mathcal{F}$ via $\ln 
\tilde{N}(u)\equiv -\beta \mathcal{F}$ with $\beta $ being subsequently
identified with the inverse temperature (provided that the system of units
is used in which the Boltzmann's constant, $k_{B}=1)$. In the present case
the role of temperature is played by $\tau .$Using these definitions, one \
can define an "energy" \ $U=<E>$ in a familiar way as $U=-\frac{\partial }{%
\partial \beta }\ln \tilde{N}(u)\equiv \tau \frac{\partial }{\partial \tau }%
\ln \tilde{N}(u)$.\ In our case, explicitly, we obtain:%
\begin{equation}
U=\tau ^{2}\frac{\partial }{\partial \tau }[\tint\nolimits_{M}u\ln
udV-N_{flat}]=\tint\nolimits_{M}\tau ^{2}\frac{\partial }{\partial \tau }%
(u\ln udV)+\frac{d}{2}\tau \tint\nolimits_{M}udV.  \tag{7.34a}
\end{equation}%
\footnote{%
Here we took into account that $\tint\nolimits_{M}udV=1$ as required.}To
proceed, we need to employ Eq.s(7.22a) and (7.22b). This leads us to 
\begin{equation}
U=\tau ^{2}\frac{\partial }{\partial \tau }[\tint\nolimits_{M}u\ln
udV-N_{flat}]=\tint\nolimits_{M}[\tau ^{2}(R+\left( \nabla f\right) ^{2})+%
\frac{d}{2}\tau ]udV.  \tag{7.34b}
\end{equation}%
This result differs slightly from that obtained by Perelman. If we are
interested in calculation of the entropy, such a difference is useful.
Indeed, since thermodynamically the relationship 
\begin{equation}
\beta U-\beta \mathcal{F}=S  \tag{7.35}
\end{equation}%
determines the entropy $S$, using \ already obtained results for $U$ and $%
\mathcal{F}$ we obtain, 
\begin{equation}
S_{+}=\tint\nolimits_{M}[\tau (R+\left( \nabla f\right) ^{2})-f+d]udV. 
\tag{7.36}
\end{equation}%
This result coincides with the result for the entropy of the Ricci expanders
obtained in Ref.[61] where a considerably more cumbersome and lengthy
pathway was chosen in order to obtain it. In order to obtain the entropy for
the Ricci shrinkers, it is sufficient to change signs when taking time
derivatives. Hence, we obtain at once:%
\begin{equation}
S_{-}=\tint\nolimits_{M}[\sigma (R+\left( \nabla f\right) ^{2})+f-d]udV, 
\tag{7.37}
\end{equation}%
again in accord with Ref.[61]. At this point, one can proceed either by
computing the heat capacity $C_{v}=(\frac{\partial }{\partial T}U$ )$_{V}$
under the constant volume or repeat the arguments for the steady solitons
adapted to the present case. The last pathway is discussed in the paper by
Cao, Ref.[66]. More physically attractive, however, is to follow the logic
of Perelman and \ to take into account that\footnote{%
As is well known from statistical mechanics, 
\begin{equation*}
C_{v}=\beta ^{2}(<E^{2}>-<E>^{2}),
\end{equation*}%
where, as before, we put $k_{B}=1.$Hence, what Perelman is simply calling a
fluctuation in Section 5 of Part I of his work, Ref.[1], is actually a heat
capacity.} 
\begin{equation*}
C_{v}=\left( \frac{\partial }{\partial \beta }U\right) \frac{\partial \beta 
}{\partial \tau }=-\left( \frac{\partial ^{2}}{\partial \beta ^{2}}\ln 
\tilde{N}(u)\right) \frac{\partial \beta }{\partial \tau },
\end{equation*}%
or%
\begin{equation}
\tau ^{2}C_{v}=\frac{\partial ^{2}}{\partial \beta ^{2}}\ln \tilde{N}(u). 
\tag{7.38}
\end{equation}
A straightforward but lengthy calculation analogous to that \ used in
Eq.(7.34) finally leads to 
\begin{equation}
\tau ^{2}C_{v}=\tau ^{4}\tint\nolimits_{M}\left\vert R_{ij}+\nabla
_{i}\nabla _{j}f-\frac{1}{2\tau }g_{ij}\right\vert ^{2}dV  \tag{7.39}
\end{equation}%
in accord with earlier obtained result, Eq.(7.29), for steady soliton
(obtained from this expression in the limit $\left\vert \tau \right\vert
\rightarrow \infty ).$ The gradient shrinking (or expanding) solitons are
respectively solutions to 
\begin{equation}
R_{ij}+\nabla _{i}\nabla _{j}f-\frac{1}{2\sigma }g_{ij}=0  \tag{7.40a}
\end{equation}%
or%
\begin{equation}
R_{ij}+\nabla _{i}\nabla _{j}f+\frac{1}{2\tau }g_{ij}=0  \tag{7.40b}
\end{equation}%
equations. By analogy with Eq.(7.30) we multiply both of these equations by $%
g^{ij}$ and sum over repeated indices in order to obtain%
\begin{equation}
R-\Delta _{g}f-\frac{d}{2\sigma }=0  \tag{7.41a}
\end{equation}%
and%
\begin{equation}
R-\Delta _{g}f+\frac{d}{2\tau }=0  \tag{7.41b}
\end{equation}%
For compact manifolds $\Delta _{g}f=0$ as before. As a result, in both cases
we obtain back Eq.(4.3a). By combining these results with those which follow
from Eq.(7.31) we\ thus have rederived \ the following result by Ivey.

\textbf{Theorem 7.20.}\textit{There are no three-dimensional solitons or
breathers on a compact connected 3-manifold other than those of constant
curvature metrics.}

\textbf{Remark 7.21}. This theorem was originally proved by Ivey, Ref[67],
who was inspired by the earlier obtained result by Hamilton [16] for
surfaces. Our derivation, however, is inspired by Perelman, Ref.[1], who
uses physical arguments. Because these arguments are not presented in
sufficient detail in his papers, \ they were left unappreciated in the works
by other mathematicians [2,3,10,61-63].

\textbf{Corollary 7.22}. In view of the obtained results it appears that, at
least for compact connected 3-manifolds, one can extract the needed physical
information for the Yamabe flow from that for the Ricci flow. We refer our
readers to works by Hamilton [5,16,70] where many illuminating additional
details can be found. The obtained result provide needed support to claims
made in Remark 7.6.

\textbf{Corollary 7.23}. The generality of arguments used for reproving the
Theorem 7.20. are such that they can be applied, in principle, to manifolds
of any dimension $d\geq 2.$ Because of this, for compact manifolds, the
Euclidean dilaton gravity described by the Eq.(7.22b) for any $d\geq 2$ is
reduced to the more familiar Euclidean gravity.

\section{Discussion}

\subsection{Other physical processes whose dynamics is described by Ricci
flows}

The goal of this paper is to find some processes taking place in the real
world providing \ a physical justification for Ricci flows. In the previous
sections such processes were found, e.g. in the form of critical dynamics.
The question arises wether there are other physical processes which also can
be used for justifying othe existence of Ricci flows. Earlier, in the text,
in Ref.[65], we mentioned the dynamics of black holes. In addition, we would
like to notice that the fundamental solution, Eq.(7.32), of the heat
equation can be interpreted as an end-to-end distancedistribution function
used in polymer physics for the computation of various averages, e.g. the
mean square end-to- end distance for the flexible polymer being modelled by
a random walk \textbf{[}69]. Under such circumstances, the time $\tau $ is
interpreted as polymer's contour length. The curvature effects can be
interpreted in terms of the polymer's backbone rigidity [71], etc. \ Upon
Fourier-Laplace transform of the distribution function, Eq.(7.3.2), the
propagator for the Klein-Gordon quantum field \ can be obtained [7, 72].
Hence, it is possible to provide an interpretation of the Ricci flow
processes in terms of some dynamics of quantum fields, etc. Whatever
processes we may discuss, it should be clear that a specific physical
process can be used only as some specific realization of the Ricci flow.
Since such situation is common for all real life processes described by the
equations of mathematical physics, the task lies in finding different
physical situations in which this flow can be realized. \ For instance,
since to conduct experiments on black holes [65] is unrealistic, one can
think of analogous processes in liquid helium as described in the monograph
by Volovik, Ref[73]. A great deal of additional condensed matter analogs of
gravity-related phenomena can be found in Refs.[74] and [75] and references
therein.

\subsection{Interplay between topology, geometry and physics}

To make our discussion complete, let us return back to the processes
involving critical dynamics studied in this paper. In view of Theorem 7.20.,
the obtained scalar curvatures either go to zero (for expanding solitons) or
blow up (for shrinking solitons) when time $t\rightarrow \infty $ [2]. \
Physically, the case of curvatures blowing up is not acceptable, however,
e.g. see Eq.(3.1) and comments related to this equation\footnote{%
It may be of some relevance to black holes, however. This topic requires
further study within the context of liquid helium and other type of
condensed matter experiments just mentioned.}. Hence, physically acceptable
manifolds must either be flat or hyperbolic in accord with Eq.(6.30) and
Remark 6.8. Although this fact was discussed in our earlier works, [47,55],
we would like to provide a sketch of relevant arguments in this subsection.

We begin with two dimensional CFT models at criticality following Ref[21%
\textbf{]}, pages 340-344, and Ref.[47], Section 5, where additional details
can be found. Many of these models can be obtained by some straightforward
modifications of the simplest Gaussian model \ defined on the torus $T^{2}$.
If $\omega _{1}$ and $\omega _{2}$ are two toral periods, one can define
their ratio $\tau $ as $\tau =\frac{\omega _{2}}{\omega _{1}}$. The number $%
\tau $ is necessarily complex with Im$\tau =\left\vert \frac{1}{\omega _{1}}%
\right\vert ^{2}\equiv y$ . In terms of these notations, the dimensionless
free energy \ $\mathcal{F}$ for the Gaussian model is obtained as follows: 
\begin{equation}
\mathcal{F=}\frac{1}{2}\ln y\left\vert \text{ }\eta (\tau )\right\vert ^{2},
\tag{8.1}
\end{equation}%
where the function $\eta (\tau )$ is the Dedekind eta function. Such a
function is known as the modular function for the once punctured torus [76%
\textbf{].}Thus, even though originally the torus topology for the Gaussian
model was used, the actual computations for this model involving the use of
the first Kronecker limit formula [47\textbf{]} leads to the final
manifestly modular invariant result for $\mathcal{F}$. The price for modular
invariance is the switch in topology: from that for a flat torus (before the
Kronecker limit is taken) to that for a punctured torus (after the limit is
taken). The punctured torus can be obtained from some parallelogram of
periods in the complex plane $\mathbf{C}$\textbf{\ }whose vertices are%
\textbf{\ }removed and whose sides are identified pair-wise\textbf{.} Since
Euler's characteristic $\chi $ for torus is $0$ while that for the punctured
torus is $-1,$the punctured torus represents an example of the hyperbolic
surface (actually an orbifold [77]) as discussed in great detail in our
work, Ref.[78]. Such a surface is not compact, however, while the results of
the previous section were developed for compact surfaces.\footnote{%
This restriction is not essential however. It is possible to extend the
theory of Ricci \ and Yamabe flows to the non compact surfaces \textbf{[}%
1,2]. Such an extension requires use of more sophisticated mathematical
methods than those used in our paper.} \ In two dimensions the situation can
be repaired by using the Schottky double construction widely used in
designing of CFT models [79]. Use of this construction causes us to replace
the punctured torus by the double torus, i.e. by the Riemann surface of
genus 2 which is also hyperbolic.

As we had demonstrated in [47] (see also Remark 6.6.) even if one begins
with the standard Euclidean cube (or any parallelepiped for that matter
whose sides are pair-wise identified thus making it a $T^{3})$ use of the
analog of the first Kronecker limit formula in 3d leads to a result similar
to Eq.(8.1) indicating that in complete accordance with 2d case, the
limiting manifold/orbifold is hyperbolic. This fact is consistent with
results of our earlier work Ref.[55] on AdS-CFT correspondence where totally
independent arguments were used to arrive at such a conclusion.

We conclude this work with a brief discussion of how one can actually design
compact hyperbolic 3 manifolds. A typical 3d hyperbolic manifold in which a
3d\ CFT lives and evolves can be built \ using a 2d punctured torus. To do
so, we would like to use the results of our earlier work, Ref.[78], in which
we noticed that such a punctured torus is the Seifert surface for the figure
eight (hyperbolic) knot. The evolution of surface automorphisms in
fictitious time create a 3 manifold. A specific hyperbolic 3 manifold known
as a 3-manifold fibered over the circle ( the puncture on the torus can be
opened up so that it is homeomorphic to a circle $S^{1}$) as discussed in
our work, Ref.[78]. The 3 manifold created in such a way will be cusped [55]
and, hence, noncompact. To make such a manifold compact one can also use a
hyperbolic double (analogous to a Schottky double as described in detail in
Ref.[80]. Obviously, the same is achieved if one starts with a Schottky
double torus and fibers it over the circle [80]. The direct 3d analog of the
two dimensional punctured torus can be constructed as a product of the trice
punctured sphere and an open interval (0,1) [80]. The obtained manifold is
homeomorphic to a cube whose upper and lower faces along with all edges are
removed. Such \ a designed 3 manifold is hyperbolic but unfortunately non
compact. Clearly, Perelman's work is not limited to compact manifolds and,
hence, his results still can be used. The only trouble with noncompactness
lies in the fact that the Yamabe flow considered in our work \ leads to the
Einstein-type spaces so that compactness is synonymous with being of
Einstein type. We mentioned this already Ref.s [26,27] in which Yamabe
functional for manifolds with boundary were discussed. More relevant to
physical applications is the work by Mazzeo et al [81] and, also Ref.[82],
in which the Yamabe problem was studied for noncompact manifolds. It would
be very interesting and challenging to study the Yamabe flow for noncompact
manifolds discussed in Refs[81,82]. Many details of the construction of
3-manifolds and orbifolds are discussed in the exceptionally well written
Ref.[77]. We conclude our paper by urging our readers to read this
reference, which is also helpful for developing a solid understanding of the
Poincare$^{\prime }$ and geometrization conjectures.

\textbf{Note added in proof.} After this work was completed several
important papers \ supporting and clarifying \ the results of our Section
8.2. came to our attention. In particular, in the paper by Long and Reid
(Algebraic and Geometric Topology 2 (2002) 285-296) it is shown that all
flat manifolds can be looked upon as cusps of hyperbolic orbifolds, e.g. see
our works, Ref.s[55],[78] for a quick introduction into cusps and Ref.[77]
for introduction to hyperbolic orbifolds. Many additional details
elaborating on the work by Long and Reid can be found in the PhD thesis by
David Ben McReinolds (arxiv:math.GT/0606571). Some related material can also
be found in the paper by X.Dai (arxiv:math.DG/0106172).

\pagebreak

\bigskip

\textbf{References}

[1] G.Perelman, The entropy formula for the Ricci flow and its geometric

\ \ \ \ \ applications, arxiv:math.DG/0211159; ibid Ricci flow with surgery
on

\ \ \ \ \ three-manifolds, arxiv:math.DG/0303109; ibid Finite extinction time

\ \ \ \ \ for the solutions to the Ricci flow on certain three- manifolds,

\ \ \ \ \ arxiv:math.DG/0307245.

[2] \ H-D.Cao, X-P.Zhu, A complete proof of the Poincare$^{\prime }$ and

\ \ \ \ \ geometrization conjectures-Application of the Hamilton-Perelman

\ \ \ \ \ theory of the Ricci flow, Asian Journal of Math.10 (2006) 165-492.

[3] J.Morgan, G.Tian, Ricci flow and the Poincare$^{\prime }$ conjecture,

\ \ \ \ arxiv:math.DG/0607607.

[4] D.Friedan, Nonlinear models in 2+$\varepsilon $ dimensions,

\ \ \ \ Ann.Phys.163 (1985) 318-419.

[5] R.Hamilton, Three-manifolds with positive Ricci curvature,

\ \ \ \ J.Diff.Geom. 17 (1982) 255-306.

[6] B.Chow, P.Lu, Lei Ni, Hamilton's Ricci Flow,

\ \ \ \ AMS Publishers, Providence, RI, 2006.

[7] A.Polyakov, Gauge Fields and Strings,

\ \ \ \ Harwood Academic Publ, New York, NY, 1987.

[8] M.Green, J.Schwarz, E.Witten, Superstring Theory, Vol.1.,

\ \ \ \ Cambridge U.Press, Cambridge,UK, 1987.

[9] D.Gross, J.Harvey, E.Martinec, R.Rohm, Heterotic string theory,

\ \ \ \ Nucl.Phys.B267 (1986) 75-124.

[10] B.Kleiner, J.Lott, Notes on Perelman's papers, arxiv: math.DG./

\ \ \ \ \ 06056667.

[11] L.Bessieres, Conjecture de Poincare$^{\prime }$ : la preuve de
R.Hamilton et

\ \ \ \ \ \ G.Perelman, SMF-Gazette 106 (2005) 7-35.

[12] M.Kapovich, Geometrization conjecture and the Ricci flow,

\ \ \ \ \ \ www.math.ucdavis.edu/\symbol{126}kapovich/EPR/ricci.ps

[13] B.Hatfield, Quantum Field Theory of Point Particles ans Strings,

\ \ \ \ \ Addison-Wesley Publ.Co, New York, NY, 1992.

[14] E.Abdalla, M.Abdalla, D.Dalmazi, A.Zadra, 2D-Gravity

\ \ \ \ \ \ in Non-Critical Strings, Springer-Verlag, Berrlin,1994.

[15] Yu.Nakayama, Liouville field theory- A decade after revolution,

\ \ \ \ \ \ arxiv:hep-th/0402009

[16] R.Hamilton, The Ricci flow on surfaces,

\ \ \ \ \ Contemp.Math.71 (1988) 237-262.

[17] D.Amit, Field Theory, the Renormalization Group

\ \ \ \ \ \ and Critical Phenomena, McGraw-Hill Inc., London, 1978.

[18] C.Itzykson, J-B.Zuber, Quantum Field Theory,

\ \ \ \ \ \ McGraw-Hill, London, 1980.

[19]\ J.Lee, T.Parker, The Yamabe problem, BAMS 17\textbf{\ }(1987), 37-91.

[20] K.Richardson, Critical points of the determinant of

\ \ \ \ \ \ the Laplace operator, J.Funct. Anal. 122 (1994), 52-83.

[21] P.Di Francesco, P.Mathieu, D.Senechal, Conformal Field Theory\textit{,}

\ \ \ \ \ \ Springer, Berlin, 1997.

[22] R.Wald, General Relativity\textit{,} The University

\ \ \ \ \ \ of Chicago Press, Chicago,1984.

[23] A.Kholodenko, K.Freed, Theta point (\textquotedblleft
tricritical\textquotedblright ) region behavior

\ \ \ \ \ \ for a polymer chain: Transition to collapse, J.Chem.Phys.

\ \ \ \ \ \ 80 (1984)900-924.

[24] L.D. Landau, E.M.Lifshitz, Statistical Physics.Part 1.Course in

\ \ \ \ \ \ Theoretical Physics. Vol.5., Pergamon Press, Oxford, 1982.

[25] H. Bray, A. Neves, Classification of prime 3-manifolds

\ \ \ \ \ \ with $\sigma -$invariant greater than RP$^{3},$

\ \ \ \ \ \ Ann.Math. 159 (2004), 407-424.

[26] J.Escobar, Conformal deformation of a Riemannian metric

\ \ \ \ \ \ to a scalar flat metric with constant mean curvature

\ \ \ \ \ \ on the boundary, Ann.Math. 136 (1992),1-50.

[27] K.Akutagawa, B. Botvinnik, Yamabe metrics on cylindrical

\ \ \ \ \ \ \ manifolds, GAFA 13 (2003), 259-333.

[28] R.M. Schoen, Variational theory for the total scalar

\ \ \ \ \ \ curvature \ functional for Riemannian metrics and related

\ \ \ \ \ \ topics, LNM 1365 (1989), 120-184.

[29] T. Aubin, \textit{\ }Some Nonlinear Problems in Riemannian Geometry,

\ \ \ \ \ \ Springer, Berlin, 1998.

[30] P.Dirac, General Theory of Relativity\textit{, }

\ \ \ \ \ \ Princeton U.Press, Princeton, 1996.

[31] H. Yamabe, On a deformation of a Riemannian structures on

\ \ \ \ \ \ compact manifolds, Osaka Math. J. 12 (1960), 21-37.

[32] A.Kholodenko, Quantum gravity and QCD in the light of works by

\ \ \ \ \ \ Hamilton and Perelman, in preparation.

[33] J.Schouten, Ricci-Calculus, Springer-Verlag,Berlin, 1954.

[34] K.Belov, About extrinsic geometry of hypersurfaces,

\ \ \ \ \ \ Soviet.Math.Izvestiya 104 (1971) 1-8 (in Russian).

[35] C.Fedischenko, V.Chernyshenko, About some generalization of

\ \ \ \ \ \ spaces of constant curvature, Tensor Analysis and Appl. 11

\ \ \ \ \ \ (1961) 269-276 (in Russian).

[36] J. Zinn-Justin, Quantum Field Theory and Critical Phenomena\textit{, }

\ \ \ \ \ \ Clarendon Press, Oxford, 1989.

[37] A.Polyakov, Quantum geometry of bosonic strings,

\ \ \ \ \ \ Phys.Lett.B 103 (1981), 207-210.

[38] O.Alvarez, Theory of strings with boundary,

\ \ \ \ \ \ Nucl.Phys.B 216 (1983), 125-184.

[39] W.Weisberger, Conformal invariants for determinants

\ \ \ \ \ \ \ of Laplacians on Riemann surfaces,

\ \ \ \ \ \ \ Comm.Math.Phys\textit{.} 112 (1987), 633-638.

[40] W.Weisberger, Normalization of the path integral measure

\ \ \ \ \ \ \ and coupling constants for bosonic strings,

\ \ \ \ \ \ \ Nucl.Phys. B 284\textbf{\ }(1987),171-200.

[41] B.Osgood, \ R. Phillips, P. Sarnak, Extremals of determinants

\ \ \ \ \ \ of Laplacians, J.Funct.Anal. 80 (1988), 148-211.

[42] B.Osgood, R. Phillips, P. Sarnak, Compact isospectral

\ \ \ \ \ \ sets of surfaces, J.Funct. Analysis 80 (1988), 212-234.

[43] B. Osgood, R. Phillips, P.Sarnak, Moduli space, heights and

\ \ \ \ \ \ isospectral sets of plane domains, Ann.Math. 129 (1989), 293-362.

[44] E. Onofri, On the positivity of the effective action in a theory

\ \ \ \ \ \ of random surfaces, Comm.Math.Phys. 86 (1982), 321-326.

[45] W. Beckner, Sharp Sobolev inequalities on the sphere and the

\ \ \ \ \ \ Moser-Trudinger inequality, $\mathrm{Ann.Math}$. 138 (1993),
213-242.

[46] M.Goulian, M. Li, Correlation functions in Liouville theory.

\ \ \ \ \ \ \ \textrm{PRL} 66 (1991), 2051-2055.

[47] A.Kholodenko, E.Ballard, From Ginzburg-Landau to Hilbert-

\ \ \ \ \ Einstein via Yamabe, Physica A 380 (2007) 115-162.

[48] S.Rosenberg, The Laplacian on a Riemannian Manifold\textit{,}

\ \ \ \ \ \ Cambridge Univ.Press, Cambridge, UK, 1997.

[49] A.Besse,\textrm{\ Einstein Manifolds}, Springer-Verlag, Berlin, 1987.

[50]\ T.Parker, S. Rosenberg, Invariants of conformal Laplacians,

\ \ \ \ \ \ \textrm{J.Diff.Geom}. 25 (1987), 535-557.

[51] K.Okikiolu, Critical metrics for the determinant

\ \ \ \ \ \ of the Laplacian in odd dimensions,

\ \ \ \ \ \ \textrm{Ann.Math}. 153 (2001), 471-531.

[52] P.Chui, Height of flat tori,

\ \ \ \ \ \ \textrm{AMS Proceedings}\textit{\ }125\textbf{\ }(1997), 723-730.

[53] S.Hawking, Zeta function regularization of path integrals in

\ \ \ \ \ \ curved space, Comm.Math.Phys.55 (1977), 133-148.

[54] Y. Muto, On Einstein's metrics,

\ \ \ \ \ J.Diff. Geom. 9 (1974), 521-530.

[55] A.Kholodenko, Boundary CFT, limit sets of Kleinian

\ \ \ \ \ groups and holography, J.Geom.Physics 35 (2000) 193-238.

[56] N.Koiso, Einstein metrics and complex structures,

\ \ \ \ \ \ \textrm{Inv.Math}. 73 (1983), 71-106.

[57]\ L.Landau, \ I.Khalatnikov, About anomalous sound

\ \ \ \ \ \ attenuation near critical points of the phase transitions of

\ \ \ \ \ \ second kind, Sov.Physics Doklady 96 (1954), 469-472.

[58] P.Chaikin, T.Lubensky, Principles of Condensed Matter Physics,

\ \ \ \ \ \ Cambridge U.Press, Cambridge, UK, 2000.

[59] H.Schwetlick, M. Struwe, Convergence of the Yamabe flow

\ \ \ \ \ \ for \textquotedblright large\textquotedblright\ energies, 
\textrm{J. Reine und Angew.Math}. 562 (2003), 59-100.

[60] A.Isihara, Statistical Physics, Academic Press, New York, NY, 1971.

[61] R.M\"{u}ller, Differential Harnac Inequalities and the Ricci Flow,

\ \ \ \ \ \ European Math.Soc.Publ.House, Zurich, SW, 2006.

[62] P.Topping, Lectures on the Ricci Flow,

\ \ \ \ \ \ Cambridge U.Press, Cambridge, UK, 2006.

[63] A.Borisenko, An introduction to Hamilton's and Perelman's work

\ \ \ \ \ on conjectures of Poincar$e^{\prime }$ and Thurston,

\ \ \ \ \ www-mechmath.univer.kharkov.ua/geometry/preprint/perelman7.pdf

[64] Y.Fujii, K-I.Maeda, The Scalar-Tensor Theory of Gravitation,

\ \ \ \ \ Cambridge U.Press, Cambridge, UK, 2003.

[65] M.Headrick,T.Wiseman, Ricci flow and black holes,

\ \ \ \ \ \ arxiv: hep-th/0606086

[66] H-D.Cao, Geometry of Ricci solitons,

\ \ \ \ \ Chin.Ann.Math.27B(2) (2006) 121-142.

[67] T.Ivey, Ricci solitons on compact three-manifolds,

\ \ \ \ \ \ Diff.Geometry and its Applications, 3 (1993) 301-307.

[68] R.Hamilton, Non-singular solutions of the Ricci flow on

\ \ \ \ \ \ three-manifolds, Comm.in Analysis and Geometry,

\ \ \ \ \ Vol.7 (1999) 695-729.

[69] P.De-Gennes, Scaling Concepts in Polymer Physics,

\ \ \ \ \ Cornell Univ.Press, Itaca, NY, 1979.

[70] R.Hamilton, Formation of singularities in the Ricci flow,

\ \ \ \ \ \ Surveys in Diff.Geom.2 (1995) 7-136.

[71] A.Kholodenko, Fermi-Bose transmutation: from

\ \ \ \ \ semiflexible polymers to superstrings,

\ \ \ \ \ Ann.Phys.202 (1990) 186-225.

[72] A.Kholodenko, Statistical mechanics of deformable droplets

\ \ \ \ on flat surfaces, J.Math.Phys.37 (1996) 1287-1313.

[73] G.Volovik, The Universe in a Helium Droplet,

\ \ \ \ \ \ Clarendon Press, Oxford, 2003.

[74] W.Zurek, Cosmological experiments in condensed matter systems,

\ \ \ \ \ \ Phys.Reports 276 (1996) 177-221.

[75] C.Barcelo, S.Liberati, M.Visser, Analogue gravity from field

\ \ \ \ \ theory normal modes, Class.Quantum Grav.18 (2001) 3595-3610.

[76] H.Iwaniec, Topics in Classical Authomorphic Forms,

\ \ \ \ \ \ AMS Publishers, Providence, RI, 1997.

[77] D.Cooper,C.Hodgson,S.Kerhoff, Three-Dimensional Orbifolds

\ \ \ \ \ and Cone Manifolds, Japan Publ.Trading Co., Tokyo, 2000.

[78] A.Kholodenko, Statistical mechanics of 2+1 gravity from

\ \ \ \ \ Riemann Zeta function and Alexander polynomial: exact results,

\ \ \ \ \ J.Geom.Phys.38 (2001) 81-139.

[79] C.Schweigerth, J.Fuchs, J.Walcher, Conformal field theory,

\ \ \ \ \ boundary conditions and applications to string theory,

\ \ \ \ \ arxiv: hep-th/0011109.

[80] K.Matsuzaki, M.Taniguchi, Hyperbolic Manifolds and

\ \ \ \ \ Kleinian Groups, Clarendon Press, Oxford, 1998.

[81] R.Mazzeo, D.Pollack, K.Uhlenbeck, Moduli spaces of singular

\ \ \ \ \ \ Yamabe metrics, AMS Journal 9 (1996) 303-344.

[82] A.Chouikha, Ricci curvature and singularities of constant scalar

\ \ \ \ \ curvature metrics, arxiv:math.DG/0501505.

\ \ \ \ \ 

\ \ \ \ 

\bigskip

\bigskip

\bigskip

\bigskip

\bigskip

\end{document}